\documentclass[pteplogo]{ptephy_v1}\usepackage{color}\usepackage{graphicx}

\usepackage{epstopdf}
\preprintnumber{YITP-16-115} 

\usepackage{amsmath}
\usepackage{amssymb}
\usepackage{bm}
\usepackage{epstopdf}
\usepackage{multirow}

\graphicspath{{./Figs/}}

\usepackage[normalem]{ulem}  

\newcommand{\comm}[1]{{\color[rgb]{1,0,1}{#1}}}

\renewcommand\sout{\bgroup \color{red} \ULdepth=-.5ex \ULset}
\newcommand\soutb{\bgroup \color{blue} \ULdepth=-.5ex \ULset}
\newcommand{\Comment}[1]{}

\newcommand{\gD}{g_{\mathrm{{\scriptscriptstyle D}}}}
\newcommand{\KMT}{\mathrm{KMT}}
\newcommand{\fm}{\mathrm{fm}}
\newcommand{\MeV}{\mathrm{MeV}}
\newcommand{\VEV}[1]{\langle{#1}\rangle}
\newcommand{\TKMT}{\mathcal{T}}
\newcommand{\TTKMT}{\widetilde{\mathcal{T}}}
\newcommand{\Norm}{\mathcal{N}}
\newcommand{\Anti}{\mathcal{A}}
\newcommand{\rhoB}{\rho_{{\scriptscriptstyle B}}}
\newcommand{\bra}[1]{\langle{#1}\,\mid}
\newcommand{\ket}[1]{\mid{#1}\,\rangle}
\newcommand{\ua}{\uparrow}
\newcommand{\da}{\downarrow}
\newcommand{\uu}{u_\ua}
\newcommand{\ud}{u_\da}
\newcommand{\du}{d_\ua}
\newcommand{\dd}{d_\da}
\newcommand{\su}{s_\ua}

\newcommand{\nup}{n_\ua}
\newcommand{\ndw}{n_\da}
\newcommand{\pup}{p_\ua}
\newcommand{\pdw}{p_\da}
\newcommand{\Lup}{\Lambda_\ua}
\newcommand{\Ldw}{\Lambda_\da}
\newcommand{\Smup}{\Sigma^-_\ua}
\newcommand{\Smdw}{\Sigma^-_\da}
\newcommand{\Szup}{\Sigma^0_\ua}

\newcommand{\Spup}{\Sigma^+_\ua}
\newcommand{\Spdw}{\Sigma^+_\da}
\newcommand{\Xmup}{\Xi^-_\ua}
\newcommand{\Xmdw}{\Xi^-_\da}
\newcommand{\Xzup}{\Xi^0_\ua}
\newcommand{\Xzdw}{\Xi^0_\da}

\begin{document}

\title{Three Baryon Interaction Generated
by Determinant Interaction of Quarks\footnote{Report No: YITP-16-115}}

\author{
\name{Akira Ohnishi}{1,\ast},
\name{Kouji Kashiwa}{1,\dag},
and \name{Kenji Morita}{1,\ddag}
}

\address{
\affil{2}{Yukawa Institute for Theoretical Physics, Kyoto University,
         Kyoto 606-8502, Japan}
\email{ohnishi@yukawa.kyoto-u.ac.jp},
{\dag}kouji.kashiwa@yukawa.kyoto-u.ac.jp,
{\ddag}kmorita@yukawa.kyoto-u.ac.jp,
}

\begin{abstract}%
We discuss the three-baryon interaction generated
by the determinant interaction of quarks,
known as the Kobayashi-Maskawa-'t Hooft (KMT) interaction.
The expectation value of the KMT interaction operator is calculated
in fully-antisymmetrized quark-cluster model wave functions
for one-, two- and three-octet baryon states.
The three-baryon potential from the KMT interaction is 
found to be repulsive for $NN\Lambda$ and $N\Lambda\Lambda$ systems,
while it is zero for the $NNN$ system.
The strength and range of the three-baryon potential
are found to be comparable to those for the $NNN$ three-body potential
obtained in lattice QCD simulations.
The contribution to the $\Lambda$ single particle potential in nuclear matter
is found to be
0.28 MeV and 0.73 MeV
in neutron matter and symmetric nuclear matter at normal nuclear
 density, respectively.
These repulsive forces
are not enough to solve the hyperon puzzle,
but may be measured in high-precision hyperisotope experiments.
\end{abstract}

\subjectindex{ D30, 
      D00, 
      D41 
}

\maketitle
\section{Introduction}

The discovery of two-solar-mass neutron
stars~\cite{Demorest:2010bx,Antoniadis:2013pzd}
has cast doubt on the equation of state (EOS)
based on conventional nuclear physics.
EOSs with nucleons and leptons for neutron star matter
can support neutron stars whose masses are $M \sim 2 M_\odot$,
where $M_\odot$ is the solar mass.
With hyperons ($Y$), however, the EOS generally becomes much softer
and many of proposed EOSs
cannot support $2 M_\odot$ neutron stars~\cite{Glendenning:1991es,Nishizaki:2002ih,Li:2008zzt,Ishizuka:2008gr,Tsubakihara:2009zb}.
This problem, referred to as the {\em hyperon puzzle}, has been attracting much attention.

Several mechanisms have been proposed so far to solve the hyperon puzzle.
One of the ideas is to assume that the crossover
transition~\cite{Masuda:2012kf,Masuda:2012ed}
from nuclear matter to quark matter takes place
at relatively low density, $2-3 \rho_0$,
with $\rho_0 \simeq 0.16~\fm^{-3}$ being the normal nuclear matter
density, instead of the often assumed first order deconfinement phase
transition at high density.
In this case, the transition may occur at lower density than the
onset of hyperon mixing via the weak interaction,
and then prevents the matter from softening owing to emergence of
the hyperons.
The crossover nature of the transition 
also circumvents the softening from the first order phase
transition.
This crossover picture can be examined by making use of
high-energy heavy-ion collision experiments.
Indeed, the direct flow collapse in Au+Au collisions at 
$\sqrt{s_{{\scriptscriptstyle NN}}}=11.5~\mathrm{GeV}$~\cite{Adamczyk:2014ipa}
seems to suggest the sudden softening of EOS~\cite{Nara:2016phs},
which could be caused by the first order phase transition.
Further studies including the isospin asymmetry dependence
of the phase transition order are necessary to examine
the crossover transition in dense neutron-rich matter.
Another idea is to modify the hyperon-nucleon ($YN$) interaction.
Within the flavor $\mathrm{SU}(3)$
coupling scheme but off the flavor-spin $\mathrm{SU}(6)$ couplings,
one can obtain a stiff hyperonic matter EOS~\cite{Weissenborn:2011ut}
with parameters including large $\bar{s}s$ contents in nucleons,
which however contradicts the recent lattice QCD calculations;
see Ref.~\cite{Freeman:2012ry} and references therein.
The third idea is to introduce the three-baryon ($3B$) interaction
involving hyperons
If the three-nucleon ($3N$) force is repulsive
and there exists repulsive $3B$ forces involving hyperons,
the EOS can support the $2M_\odot$ neutron stars by suppressing the hyperon mixing at high densities.

The role of the $3N$ force has been extensively discussed in nuclear physics.
Ab initio calculations with two-nucleon interactions cannot reproduce
the binding energies of three-nucleon systems ($t$ and $^3\mathrm{He}$)
and the nuclear matter saturation point, and it has been noticed that
three-nucleon ($3N$) force is necessary
to reproduce these fundamental aspects of nucleon many-body
systems~\cite{Fujita:1957zz,Pieper:2001mp,Yamamoto:2014jga,Epelbaum:2008ga,Machleidt:2011zz,Friedman:1981qw,Akmal:1998cf,Kohno:2013dsa}.
The $3N$ force contains two-pion P-wave exchange term with an
intermediate $\Delta$ excitation (Fujita-Miyazawa
force)~\cite{Fujita:1957zz} and shorter-range phenomenological
term~\cite{Pieper:2001mp}. One of the physical pictures of the
short-range $3N$ force is the multi-pomeron
exchange~\cite{Yamamoto:2014jga}.
Recently, $3N$ force has been derived systematically
in the chiral effective field theory,
where energy-independent $3N$ forces appear at the next-to-next-to-leading
order~\cite{Epelbaum:2008ga,Machleidt:2011zz}.
Microscopic calculations including these $3N$ forces show that
the $3N$ force is relevant to the nuclear matter saturation
and the nuclear matter EOS at high
densities~\cite{Pieper:2001mp,Yamamoto:2014jga,Friedman:1981qw,Akmal:1998cf,Kohno:2013dsa},
and also to deuteron-proton scattering observables~\cite{Sekiguchi:2002}.
Moreover, there haven been remarkable progress in the lattice QCD simulation for
the $3N$ force based on the Nambu-Bethe-Salpeter wave function,
i.e. HAL QCD method~\cite{Doi:2011gq}.
At present, such lattice QCD simulations have been performed for
larger $\pi$ masses than the physical $\pi$ mass and then their data
still have large error-bars. Cooperations between lattice QCD
simulations, experiments and phenomenological approaches including
present study should play an important role to understand nature of the
$3N$ force and also other $3B$ forces in future.

So far $3B$ forces involving hyperons have not been known well.
The currently available hypernuclear data are not enough
to determine the $3B$ force involving hyperons precisely
at the level of the $3N$ force,
then we have to rely on some models or assumptions.
In Ref.~\cite{Yamamoto:2014jga}, 
the repulsive $3B$ forces are assumed to work universally
for $YNN$, $YYN$, $YYY$ as well as for $NNN$
in the multi-pomeron exchange mechanism.
By fixing the multi-pomeron exchange strength
by the nuclear elastic scattering,
the nuclear matter saturation and hypernuclear separation energies
are reproduced in the $G$-matrix calculations,
and the $2M_\odot$ neutron stars are found to be supported.
In Ref.~\cite{Lonardoni:2013gta},
the authors adopt a phenomenological hyperon-nucleon 
potential,
which contains the $\Lambda{N}$ forces consistent
with the $\Lambda{p}$ scattering data
and strongly repulsive Wigner type $\Lambda NN$ force~\cite{Bodmer:1984gc}.
The $\Lambda$ separation energies are well reproduced
in the auxiliary field diffusion Monte-Carlo calculations,
and the EOS fitted to the Monte-Carlo results can support
the $2M_\odot$ neutron stars when the $\Lambda{NN}$ forces are
included~\cite{Lonardoni:2014bwa}.
In this treatment, 
the $\Lambda$ separation energies in heavy $\Lambda$ hypernuclei 
are calculated to be more than $60$ MeV with the two-body ($2B$)
forces,
while they are reduced to be less than $30$ MeV with the
$3B$ forces.
The effects of the $3B$ potential are very large,
and it is natural to deduce that the four- and more baryon potential
would be as important as the $3B$ potential.
In Ref.~\cite{Miyatsu:2013hea},
internal structure modification of baryons is taken into account
in the framework of the chiral quark meson coupling model.
The modification leads to the $\sigma$ dependence of
the baryon-$\sigma$ coupling,
or the baryon-$\sigma$-$\sigma$ multi-body coupling,
where $\sigma$ is the isoscalar scalar meson field.
Because of the multi-body coupling and the Fock term contribution,
hyperons are suppressed at high densities,
and $2M_\odot$ neutron stars can be supported within the observation errors.
While these approaches are successful in explaining the existence
of $2M_\odot$ neutron stars with EOS including hyperons,
it is preferable to derive the $3B$ interaction 
from a view point of the quark dynamics.

\begin{figure}[tbhp]
\centerline{\includegraphics[bb=100 430 180 530,clip,width=3cm]{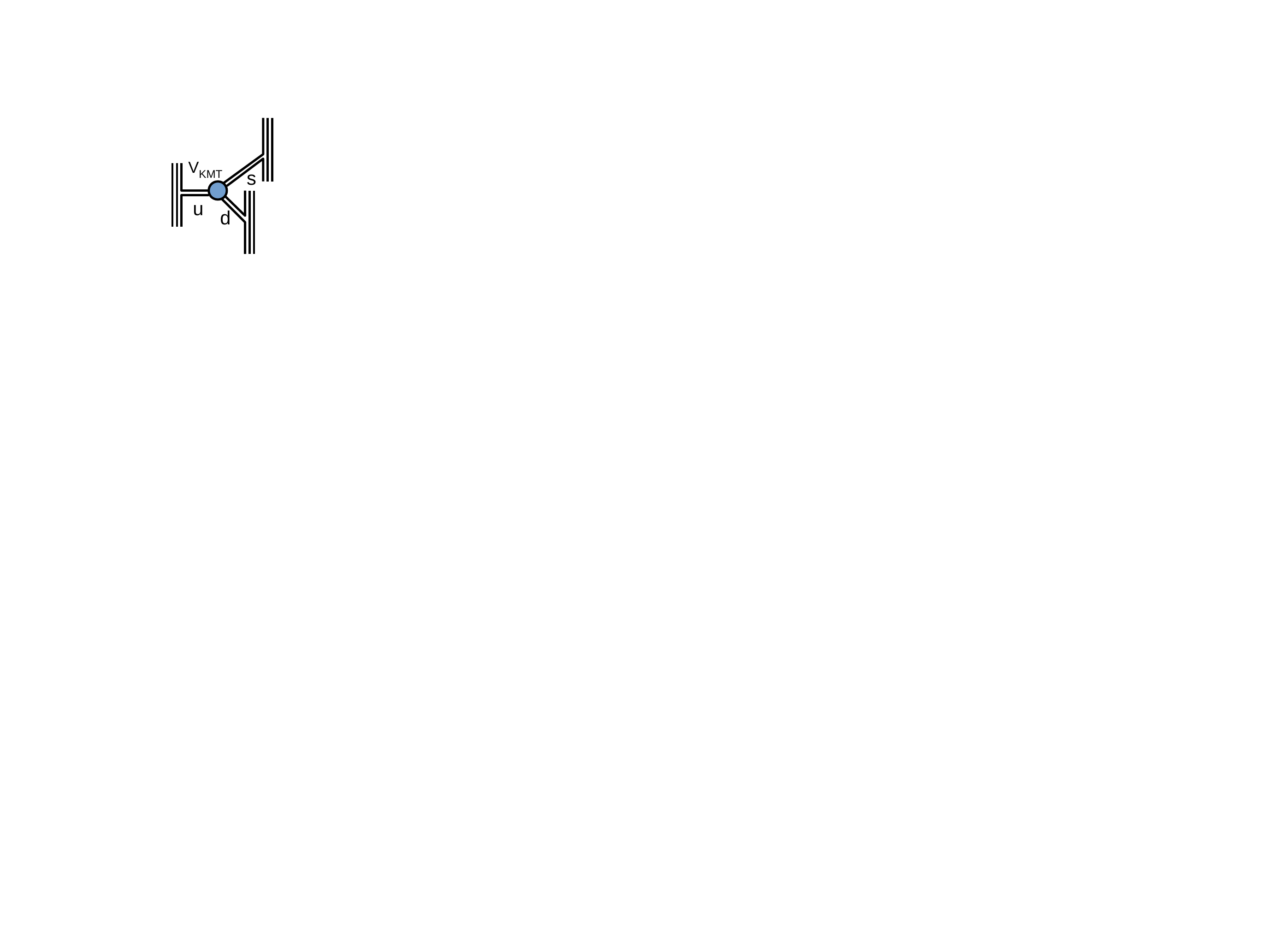}}
\caption{Schematic picture of three-baryon interaction
generated by the KMT interaction}\label{Fig:VKMT}
\end{figure}

One of the possible origins of the $3B$ interaction
is the determinant interaction of quarks,
referred to as the Kobayashi-Maskawa-'t Hooft (KMT) interaction
~\cite{Kobayashi:1970ji,Kobayashi:1971qz,'tHooft:1976fv,'tHooft:1986nc},
\begin{align}
\mathcal{L}_\KMT=&\gD\,\left(
\det \Phi + \mathrm{h.c.}
\right)
\ ,\label{Eq:KMT}\\
\Phi_{ij}=&\bar{q}_j(1-\gamma_5)q_i
\ \label{Eq:Phi},
\end{align}
where $\mathrm{det}$ denotes the determinant with respects to the
flavor indices, $i$ and $j$.
With three-flavors of quarks ($N_f=3$),
the KMT interaction replaces three quarks having different flavors
with three quarks, then it can generate the $3B$ interaction;
The 3B system having at least one $u$, $d$ and $s$ quarks
can be connected by the KMT interaction as shown
schematically in Fig.~\ref{Fig:VKMT}.

\begin{figure}[tbhp]
\centerline{\includegraphics[bb=50 410 490 520,clip,width=15cm]{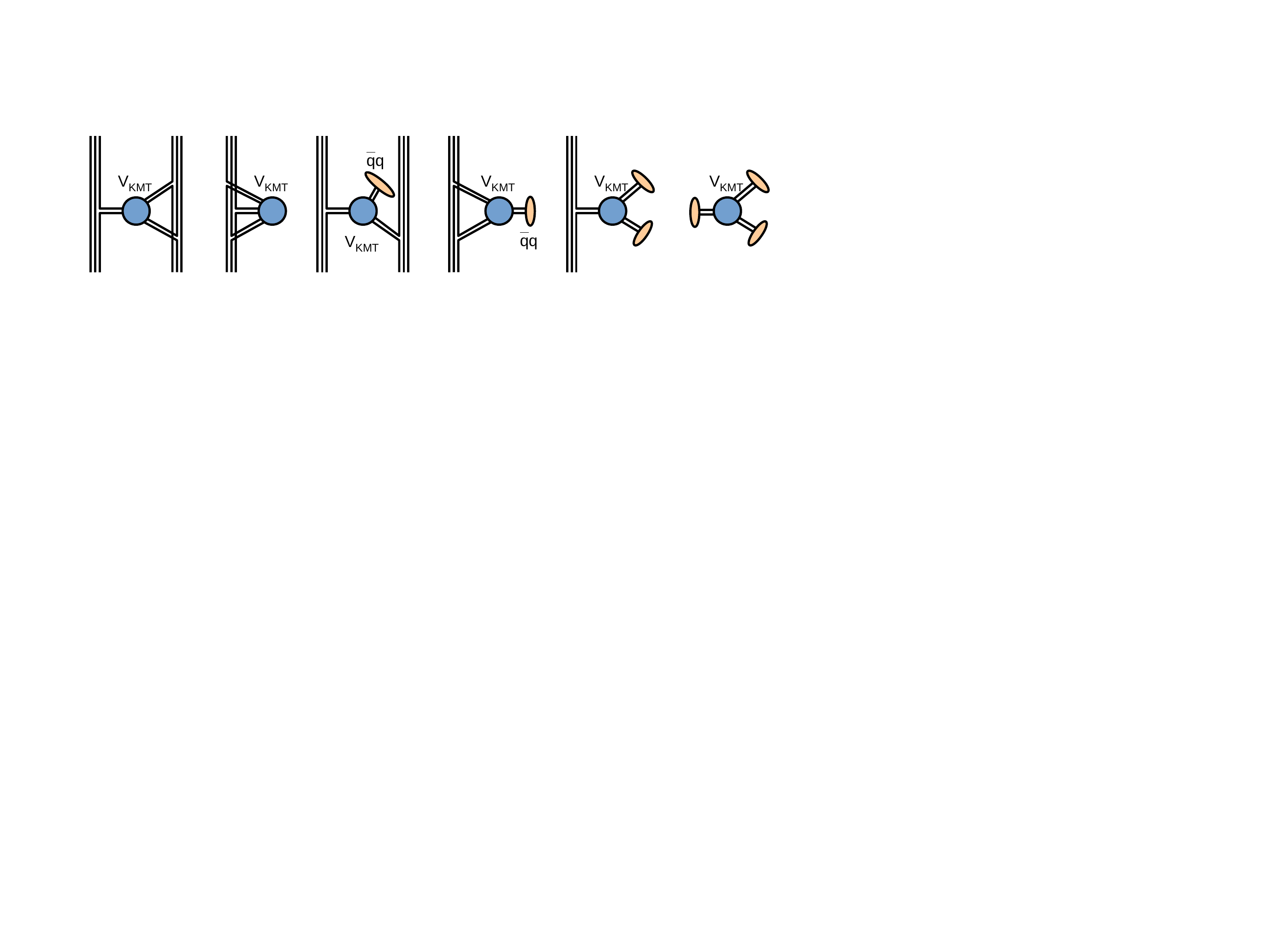}}
\caption{Other interaction terms from the KMT interaction}\label{Fig:VKMTother}
\end{figure}

The KMT interaction was introduced in order to account
for the axial $U(1)$ ($\mathrm{U}(1)_A$)
anomaly~\cite{Kobayashi:1970ji,Kobayashi:1971qz},
and it was found to be generated
by instantons~\cite{'tHooft:1976fv,'tHooft:1986nc}.
The strength of the KMT interaction $\gD$ can be fixed by
reproducing basic quantities of mesons.
For example, $\gD$ was determined to be
$\gD\Lambda^5=-9.29$~\cite{Hatsuda:1994pi}
or $\gD\Lambda^5=-12.36$~\cite{Rehberg:1995kh}
in the Nambu--Jona-Lasinio (NJL) model~\cite{Nambu:1961tp,Nambu:1961fr}
by fitting the $\pi$, $K$ and $\eta'$ masses and the pion
decay constant with other model parameters.
Since the $\mathrm{U}(1)_A$ anomaly pushes up the $\eta'$ mass
($m_{\eta^\prime}$), the value of $\gD$ is essentially fixed by
$m_{\eta^\prime}$~\cite{Hatsuda:1994pi}.

From the above considerations, the 3$B$ interaction
from the KMT interaction has favorable features to resolve the hyperon puzzle.
It acts only on systems including strange quarks,
and it is probably repulsive because of the negative value of $\gD$.
In the mean-field treatment for the NJL model,
the KMT interaction pushes up the constituent quark mass~\cite{Hatsuda:1994pi},
$M_u=m_u - 2 g_s\VEV{\bar{u}u} - 2\gD \VEV{\bar{d}d} \VEV{\bar{s}s}$,
$M_d=m_d - 2 g_s\VEV{\bar{d}d} - 2\gD \VEV{\bar{s}s} \VEV{\bar{u}u}$,
and
$M_s=m_s - 2 g_s\VEV{\bar{s}s} - 2\gD \VEV{\bar{u}u} \VEV{\bar{d}d}$,
where $m_{u,d,s}$ represents the current quark mass,
$g_s$ is the four-fermion scalar coupling,
and $\VEV{\bar{q}q} < 0$ is the quark condensate.
%
The KMT interaction is found to act repulsively
also in the $\Lambda\Lambda$ interaction~\cite{Takeuchi:1990qj}.
While the color-magnetic interaction is strongly attractive
in the flavor singlet six-quark state, the $H$ dibaryon~\cite{Jaffe:1976yi},
the KMT interaction acts repulsively and the mass of
$H$ may be pushed up above the $\Lambda\Lambda$ threshold.

In this article, we discuss the 3$B$ interaction
generated by the KMT interaction in the quark cluster model.
We adopt here the non-relativistic treatment for the wave functions of quarks
and the KMT interaction.
We consider the non-relativistic wave functions of quarks in baryons,
and the spatial wave function is assumed to be $(0s)^3$;
all quarks are assumed to be in the $s$-wave state of the harmonic
oscillator potential and to have the same spatial extension.
In the non-relativistic treatment of the KMT interaction, we ignore the
$\gamma_5$ part in $\Phi_{ij}$, then the KMT interaction
becomes the zero-range three-body interaction among quarks (six-fermi interaction),
which modifies the flavor but does not modify the spin and color of quarks.

This paper is organized as follows.
In Sec. \ref{Sec:WF},
we evaluate the norm of the one-, two-, and three-baryon
wave functions in the quark cluster model.
In Sec.~\ref{Sec:KMT}, we quantify the 3$B$
potential generated by the KMT interaction based on
the three-baryon wave function and discuss its implication to the
hyperon puzzle. Section \ref{sec:summary} is devoted to the summary. 

\section{Quark cluster model wave function}
\label{Sec:WF}

\subsection{One baryon state}

We consider here the flavor-spin SU(6) wave function of octet baryons.
For example, 
the color, flavor, spin wave function for $\nup$, $\ndw$ and $\Lup$
are given as
\begin{align}
\psi(\nup)
=&\frac{1}{\sqrt{3!}}\,\frac{1}{2\sqrt{3}}\,
\varepsilon_{abc}\left[\du\dd\uu+\dd\du\uu-2\du\du\ud\right]^{abc}
\ ,\\
\psi(\ndw)
=&\frac{1}{\sqrt{3!}}\,\frac{1}{2\sqrt{3}}\,
\varepsilon_{abc}\left[\dd\du\ud+\du\dd\ud-2\dd\dd\uu\right]^{abc}
\ ,\\
\psi(\Lup)
=&\frac{1}{\sqrt{3!}}\,\frac{1}{\sqrt{2}}\,
\varepsilon_{abc}\left[\uu\dd\su-\ud\du\su\right]^{abc}
\ ,
\end{align}
where $\varepsilon_{abc}$ is the Levi-Civita tensor with respect
to color indices.
The first $1/\sqrt{3!}$ factor is the color symmetry factor, which is common
to all baryons.
The normalization factor is determined to normalize the fully antisymmetrized
wave function, $\psi_\mathcal{A}=\mathcal{A}/\sqrt{3!}\times\psi$.
The antisymmetrizer $\mathcal{A}$ is defined as
\begin{align}
\mathcal{A}[q_1 q_2 \cdots q_n]
=\sum_P \mathrm{sgn}(P)\, q_{P(1)} q_{P(2)} \cdots q_{P(n)}
\ ,
\end{align}
where 
$P$ denotes permutation, $q_i$ is the quark wave function, 
and $\mathrm{sgn}(P)$ is the sign (or signature) of the permutation $P$.
The flavor and spin wave functions for octet baryons are summarized
in Table \ref{Table:Octet} for completeness.

\begin{table}
\caption{Octet baryon wave functions with the spin-up state.
The fully antisymmetrized wave function is given as
$\ket{\psi_\mathcal{A}}=\mathcal{A}/\sqrt{3!}\times\varepsilon_{abc}/\sqrt{3!}\times\left[\mid\mathrm{Flavor}\,\rangle\otimes\mid\mathrm{Spin}\,\rangle\otimes\mid\mathrm{Spatial~w.f.}\rangle\right]^{abc}$.
}\label{Table:Octet}
\centerline{
\begin{tabular}{llc}
\hline
\hline
$B$ & \multicolumn{1}{c}{$\mid\mathrm{Flavor}\rangle$} & $\mid\mathrm{Spin}\rangle$ \\
\hline
$\nup$  &$\ket{ddu}/\sqrt{2}$ & $\ket{\ua\da\ua+\da\ua\ua-2\ua\ua\da}/\sqrt{6}$ \\
$\pup$  &$\ket{uud}/\sqrt{2}$ & $\ket{\ua\da\ua+\da\ua\ua-2\ua\ua\da}/\sqrt{6}$ \\
$\Lup$  &$\ket{uds}$          & $\ket{\ua\da\ua-\da\ua\ua}/\sqrt{2}$ \\
$\Smup$ &$\ket{dds}/\sqrt{2}$ & $\ket{\ua\da\ua+\da\ua\ua-2\ua\ua\da}/\sqrt{6}$ \\
$\Szup$ &$\ket{uds}$          & $\ket{\ua\da\ua+\da\ua\ua-2\ua\ua\da}/\sqrt{6}$ \\
$\Spup$ &$\ket{uus}/\sqrt{2}$ & $\ket{\ua\da\ua+\da\ua\ua-2\ua\ua\da}/\sqrt{6}$ \\
$\Xmup$ &$\ket{ssd}/\sqrt{2}$ & $\ket{\ua\da\ua+\da\ua\ua-2\ua\ua\da}/\sqrt{6}$ \\
$\Xzup$ &$\ket{ssu}/\sqrt{2}$ & $\ket{\ua\da\ua+\da\ua\ua-2\ua\ua\da}/\sqrt{6}$ \\
\hline
\hline
\end{tabular}}
\end{table}

The norm of the one baryon state is obtained as 
\begin{align}
\mathcal{N}_\mathcal{A}\equiv
&\langle{\psi_\mathcal{A}(\nup)}\mid{\psi_\mathcal{A}(\nup)}\rangle
=\frac{1}{3!}
\langle{\mathcal{A}\,[\psi(\nup)]}
\mid
{\mathcal{A}\,[\psi(\nup)]}\rangle
=
\VEV{\psi(\nup)\mid\mathcal{A}\,[\psi(\nup)]}
\nonumber\\
=&\frac{1}{3!}\sum_{i,j} c^*_i c_j
\langle{\varepsilon_{abc}\phi_i^{abc}(\nup)}\mid
{\mathcal{A}\,[\varepsilon_{def}\phi_j^{def}(\nup)]}\rangle
\nonumber\\
=&\sum_{i,j} c^*_i c_j
\langle{\phi_i(\nup)}\mid
{[\mathcal{S}_\mathrm{fss}\,\phi_j(\nup)]}\rangle_\mathrm{fss}
=\sum_{i,j} c^*_i c_j\,\sum_P F_P(\phi_i,\phi_j)
\ ,
\end{align}
where
$c_i=1/2\sqrt{3}, 1/2\sqrt{3}, -1/\sqrt{3}\ (i=1,2,3)$
and
$\phi_i=\du\dd\uu, \dd\du\uu, \du\du\ud\ (i=1,2,3)$
are the flavor-spin coefficient and component of the one-baryon wave function,
respectively.
The antisymmetrizer $\mathcal{A}$ acts on the color,
flavor, spin, and spatial coordinate wave functions,
while the symmetrizer in the flavor-spin-spatial (fss)
coordinates, $\mathcal{S}_\mathrm{fss}$,
does not exchange color indices.
In the last line, the matrix element is obtained
in the flavor, spin, and spatial coordinate wave functions, 
$F_P(\phi_i,\phi_j)=\langle\phi_i\mid P\phi_j\rangle_\mathrm{fss}$.
For example, one of the elements looks like
\begin{align}
&\langle{\phi_1(\nup)}\mid
{[\mathcal{S}_\mathrm{fss}\,\phi_2(\nup)]}\rangle_\mathrm{fss}
= \sum_P F_P(\phi_1,\phi_2)
\nonumber\\
&=\langle{\du\dd\uu}
\mid{\dd\du\uu+\du\uu\dd+\uu\dd\du+\dd\uu\du+\uu\du\dd+\du\dd\uu}\rangle_\mathrm{fss}
=1
\ ,
\end{align}
provided that the spatial wave function is common to all quarks and normalized.

\subsection{Two baryon states}

Two-baryon states are defined as the product of two\comm{-}baryon wave functions
fully antisymmetrized in color, flavor, spin and spatial coordinates.
We show the case of $\nup\ndw$, as an example,
\begin{align}
\mid{\psi_\mathcal{A}(\nup,\ndw)}\rangle
=& \frac{1}{\sqrt{6!}}\mid\mathcal{A}[\psi(\nup)\psi(\ndw)]\rangle
\ .
\end{align}
The norm of the above two-baryon state is obtained as
\begin{align}
\mathcal{N}_\mathcal{A}=
&\langle{\psi_\mathcal{A}(\nup,\ndw)}
\mid{\psi_\mathcal{A}(\nup,\ndw)}\rangle
\nonumber\\
=&\langle\psi(\nup)\psi(\ndw)\mid
\mathcal{A}[\psi(\nup)\psi(\ndw)]\rangle
\nonumber\\
=&\frac{1}{(3!)^2}\sum_{i,j,k,l}
c^*_i(\nup) c^*_j(\ndw)
c_k(\nup) c_l(\ndw)
\,\varepsilon_{abc}
\,\varepsilon_{def}
\,\varepsilon_{a'b'c'}
\,\varepsilon_{d'e'f'}
\nonumber\\
&\times
\langle
\phi_i^{abc}(\nup)\phi_j^{def}(\ndw)
\mid
\mathcal{A}\,[\phi_k^{a'b'c'}(\nup)\phi_l^{d'e'f'}(\ndw)]
\rangle
\nonumber\\
=&\sum_{i,j,k,l} c^*_i(\nup) c^*_j(\ndw) c_k(\nup) c_l(\ndw)
\sum_P C_P(\phi_i\phi_j,\phi_k\phi_l)\,F_P(\phi_i\phi_j,\phi_k\phi_l)
\ ,
\end{align}
The color and flavor-spin-spatial factors are shown by
$C_P$ and $F_P$.
The exchange of quarks among baryons gives rise to a sign factor,
which cannot be absorbed in the Levi-Civita tensor $\varepsilon$,
\begin{align}
\mathcal{A}[1^a1^b1^c2^d2^e2^f]
=1^a1^b1^c2^d2^e2^f
-1^a1^b2^d1^c2^e2^f
+1^a2^e2^d1^c1^b2^f
+\cdots
\ ,\label{Eq:ExchExample}
\end{align}
where $1^a$ ($2^d$) shows one of the quarks in the first (second) baryon
having the color index $a$ ($d$).
The color factor is obtained as the contraction of $\varepsilon$.
For the case of the second term in Eq.~\eqref{Eq:ExchExample},
the first (second) three colors in the bra state need to be $abd$ ($cef$)
to have a finite matrix element,
and we obtain
\begin{align}
C_P
=-\frac{1}{(3!)^2}
\varepsilon_{abd}\varepsilon_{cef}
\varepsilon_{abc}\varepsilon_{def}
=-\frac{1}{36}\,
 2\delta_{dc}\,2\delta_{cd}=-\frac{1}{3}
\ .
\end{align}
The color factor in the two-baryon case is found to be
$C_P=1, -1/3, 1/3, -1$ for zero, one, two and three quark
exchanges between baryons, respectively.
The flavor-spin-spatial factor for the permutation $P$ is
\begin{align}
F_P(\phi_i\phi_j,\phi_k\phi_l)
=\langle\phi_i(\nup)\phi_j(\ndw)\mid
P[\phi_k(\nup)\phi_l(\ndw)]\rangle_\mathrm{fss}
=0~\mathrm{or}~1
\ ,
\end{align}
provided that two baryons are located at the same spatial point.

\subsection{Three baryon states}

Three-baryon states are defined in a way similar to
the two-baryon states.
For the $\nup\ndw\Lup$ state, the three-baryon wave function reads
\begin{align}
\mid{\psi_\mathcal{A}(\nup,\ndw,\Lup)}\rangle
=& \frac{1}{\sqrt{9!}}\mid\mathcal{A}
[\psi(\nup)\psi(\ndw)\psi(\Lup)]\rangle
\ .
\end{align}
The norm of three-baryon states are obtained as
\begin{align}
\mathcal{N}_\mathcal{A}=
&\langle{\psi_\mathcal{A}(\nup,\ndw,\Lup)}
\mid{\psi_\mathcal{A}(\nup,\ndw,\Lup)}\rangle
\nonumber\\
=&\langle\psi(\nup)\psi(\ndw)\psi(\Lup)\mid
\mathcal{A}[\psi(\nup)\psi(\ndw)\psi(\Lup)]\rangle
\nonumber\\
=&\frac{1}{(3!)^3}\sum_{i,j,k,l,m,n}
c^*_{ijk}(\nup\ndw\Lup) c_{lmn}(\nup\ndw\Lup)
\,\varepsilon_{abc}
\,\varepsilon_{def}
\,\varepsilon_{ghi}
\,\varepsilon_{a'b'c'}
\,\varepsilon_{d'e'f'}
\,\varepsilon_{g'h'i'}
\nonumber\\
&\times
\langle
\phi_i^{abc}(\nup)\phi_j^{def}(\ndw)\phi_k^{ghi}(\Lup)
\mid
\mathcal{A}\,[\phi_l^{a'b'c'}(\nup)\phi_m^{d'e'f'}(\ndw)\phi_n^{g'h'i'}(\Lup)]
\rangle
\nonumber\\
=&\sum_{i,j,k,l,m,n} 
c^*_{ijk}(\nup\ndw\Lup) c_{lmn}(\nup\ndw\Lup)
\sum_P C_P(\phi_{ijk},\phi_{lmn})\,F_P(\phi_{ijk},\phi_{lmn})
\ ,
\end{align}
where 
$c_{ijk}(B_1B_2B_3)=c_i(B_1)c_j(B_2)c_k(B_3)$,
$\phi_{ijk}=\phi_i\phi_j\phi_k$,
and $C_P$ and $F_P$ are the color and flavor-spin-spatial
factors, respectively.

\begin{figure}[tbhp]
\includegraphics[bb=100 40 750 560,width=14cm]{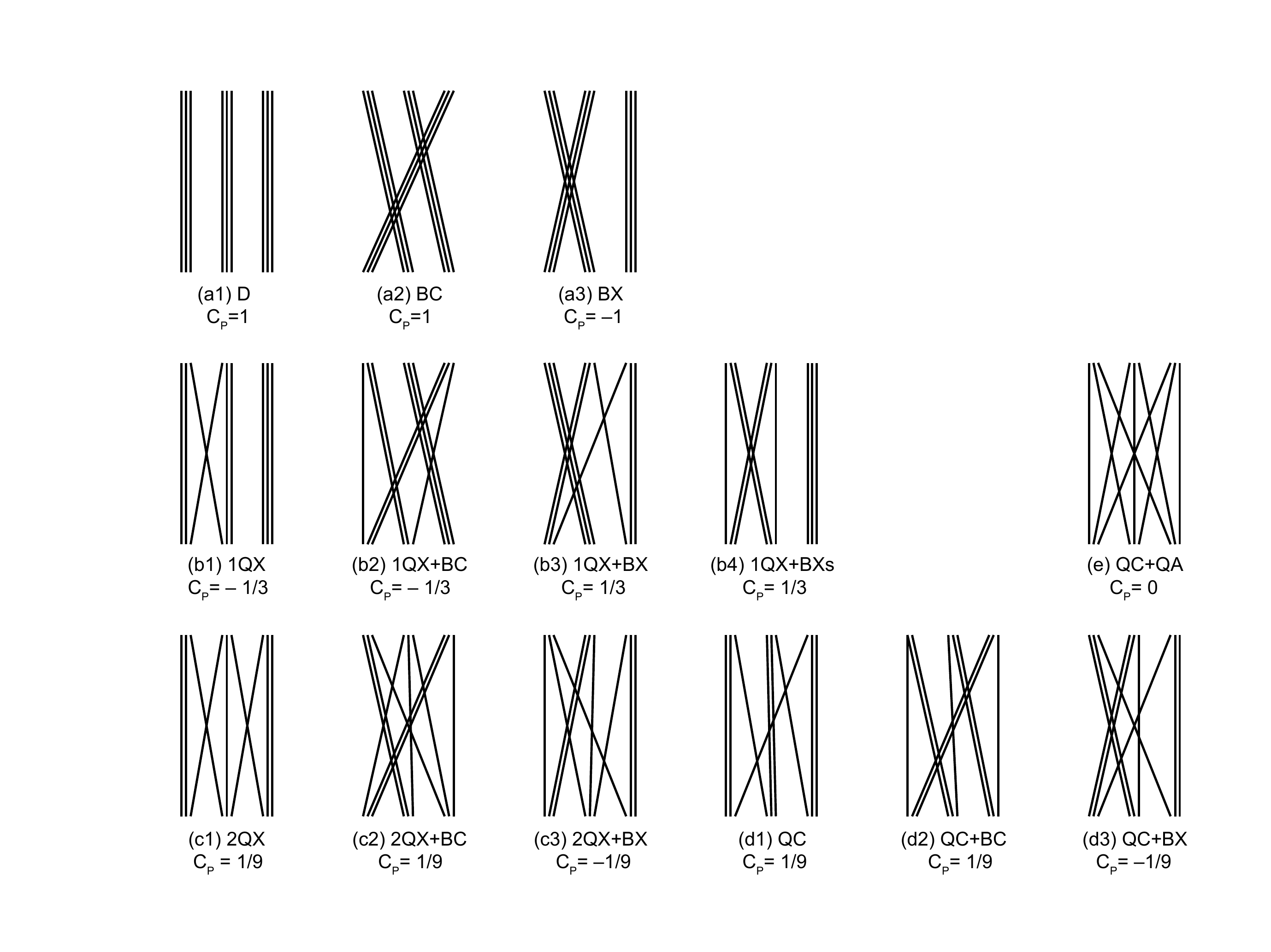}
\caption{Quark exchange diagrams and the corresponding color factor.
See the text for the description of the quark exchanges.
}\label{Fig:ColorTypes}
\end{figure}

Calculation of the color factor for the three-baryon states is
straightforward but lengthy.
We show one of the non-trivial examples,
\begin{align}
&-\frac{1}{(3!)^3}
\,\varepsilon_{abc}
\,\varepsilon_{def}
\,\varepsilon_{ghi}
\,\varepsilon_{a'b'c'}
\,\varepsilon_{d'e'f'}
\,\varepsilon_{g'h'i'}
\langle
1^a1^b1^c2^d2^e2^f3^g3^h3^i
\mid
1^{a'}1^{b'}3^{g'}2^{d'}3^{h'}3^{i'}1^{c'}2^{e'}2^{f'}
\rangle
\nonumber\\
=&-\frac{1}{(3!)^3}
\,\varepsilon_{abc}
\,\varepsilon_{def}
\,\varepsilon_{ghi}
\,\varepsilon_{abg}
\,\varepsilon_{dhi}
\,\varepsilon_{cef}
\langle 111222333 \mid 113233122 \rangle_\mathrm{fss}
\nonumber\\
=&-\frac{1}{9}
\langle 111222333 \mid 113233122 \rangle_\mathrm{fss}
\ .
\end{align}
In this case the color factor is $C_P=-1/9$.

Exchanges of quarks among three baryons
can be categorized into the following types.
\begin{description}
 \item[(a1)] Direct term (No quark exchange) (D).
 \item[(a2)] Cyclic baryon exchange (BC).
 \item[(a3)] Baryon exchange (BX).
 \item[(b1)] One quark pair exchange between two baryons (1QX).
 \item[(b2)] One quark pair exchange between two baryons after cyclic
	    baryon exchange (1QX+BC).
 \item[(b3)] One quark pair exchange between two baryons after baryon
	    exchange of the different baryon pair (1QX+BX).
 \item[(b4)] One quark pair exchange between two baryons after baryon exchange
	    of the same baryon pair (1QX+BXs).
	    (or two quark pair exchange between two baryons).
 \item[(c1)] Two quark pair exchange among three baryons with a stand baryon
	    (2QX).
 \item[(c2)] Two quark pair exchange among three baryons with a stand baryon
	    after cyclic baryon exchange (2QX+BC).
 \item[(c3)] Two quark pair exchange among three baryons with a stand baryon
	    after baryon exchange (2QX+BX).
 \item[(d1)] Cyclic three quark exchange (QC).
 \item[(d2)] Cyclic three quark exchange after cyclic baryon exchange
	    (QC+BC).
 \item[(d3)] Cyclic three quark exchange after baryon exchange (QC+BX).
 \item[(e)] ~~Cyclic three-quark and anti-cyclic three-quark exchange among three baryons
    (QC+QA).
\end{description}
We schematically show the corresponding diagrams for these types
in Fig.~\ref{Fig:ColorTypes}.

\begin{table}
\caption{Color factor.
See the text for the quark exchange.
}\label{Table:ColorFactor}
\begin{minipage}[t]{8cm}
\begin{tabular}[t]{lcl}
\hline
\hline
Permutation & $C_P$ & Type \\
\hline
111 222 333 & 1 	&\multirow{1}{2.5cm}{D (a1)}\\
\hline
222 333 111 & 1 	&\multirow{2}{2.5cm}{BC (a2)}\\
333 111 222 & 1 	&\\
\hline
111 333 222 & $-1$ 	&\multirow{3}{2.5cm}{BX (a3)}\\
333 222 111 & $-1$ 	&\\
222 111 333 & $-1$ 	&\\
\hline
111 223 233 & $-1/3$ 	&\multirow{3}{2.5cm}{1QX (b1)}\\
113 222 133 & $-1/3$ 	& \\
112 122 333 & $-1/3$ 	& \\
\hline
133 113 222 & $-1/3$ 	&\multirow{6}{2.5cm}{1QX+BC (b2)}\\
333 112 122 & $-1/3$ 	&\\
233 111 223 & $-1/3$ 	&\\
223 233 111 & $-1/3$ 	&\\
222 133 113 & $-1/3$ 	&\\
122 333 112 & $-1/3$ 	&\\
\hline
113 133 222 & $1/3$ 	&\multirow{6}{2.5cm}{1QX+BX (b3)}\\
112 333 122 & $1/3$ 	&\\
233 223 111 & $1/3$ 	&\\
333 122 112 & $1/3$ 	&\\
223 111 233 & $1/3$ 	&\\
222 113 133 & $1/3$ 	&\\
\hline
111 233 223 & $1/3$ 	&\multirow{3}{2.5cm}{1QX+BXs(b4)}\\
133 222 113 & $1/3$ 	&\\
122 112 333 & $1/3$ 	&\\
\hline
\hline
\end{tabular}%
\end{minipage}\begin{minipage}[t]{8cm}
\begin{tabular}[t]{lcl}
\hline
\hline
Permutation & $C_P$ & Type \\
\hline
123 122 133 & $1/9$ &\multirow{3}{2.5cm}{2QX (c1)}\\
112 123 233 & $1/9$ &\\
113 223 123 & $1/9$ &\\
\hline
123 113 223 & $1/9$ &\multirow{6}{2.5cm}{2QX+BC (c2)}\\
133 123 122 & $1/9$ &\\ 
233 112 123 & $1/9$ &\\
123 233 112 & $1/9$ &\\
223 123 113 & $1/9$ &\\
122 133 123 & $1/9$ &\\
\hline
113 123 223 & $-1/9$ &\multirow{9}{2.5cm}{2QX+BX (c3)}\\
112 233 123 & $-1/9$ &\\ 
123 133 122 & $-1/9$ &\\ 
123 223 113 & $-1/9$ &\\ 
233 123 112 & $-1/9$ &\\ 
133 122 123 & $-1/9$ &\\ 
123 112 233 & $-1/9$ &\\ 
122 123 133 & $-1/9$ &\\ 
223 113 123 & $-1/9$ &\\ 
\hline
112 223 133 & $1/9$ &\multirow{2}{2.5cm}{QC (d1)}\\
113 122 233 & $1/9$ &\\
\hline
233 113 122 & $1/9$ &\multirow{4}{2.5cm}{QC+BC (d2)}\\
133 112 223 & $1/9$ &\\
122 233 113 & $1/9$ &\\
223 133 112 & $1/9$ &\\
\hline
112 133 223 & $-1/9$ &\multirow{6}{2.5cm}{QC+BX (d3)}\\
113 233 122 & $-1/9$ &\\
133 223 112 & $-1/9$ &\\
233 122 113 & $-1/9$ &\\
223 112 133 & $-1/9$ &\\
122 113 233 & $-1/9$ &\\
\hline
123 123 123 & $0$ &\multirow{1}{2.5cm}{QC+QA (e)}\\
\hline
\hline
\end{tabular}
\end{minipage}
\end{table}

We find that the color factor is $\pm1$ when there is no quark exchange
among baryons (a1, a2, a3).
When one baryon keeps its three quarks (b1-b4),
the color factor is $\pm 1/3$.
In the case where all three baryons are involved in quark exchanges,
the color factor is $\pm 1/9$ in most of the cases (c1-c3, d1-d3).
The single exception is (e), $C_P=0$, where the ket state is made of
three baryons, each of which contains one quark from each baryon 
in the bra state.
We summarize the color factor in Table~\ref{Table:ColorFactor},
and the norm in some of  two- and three-baryon systems
in Table~\ref{Table:KMT}.

The norm of the three-baryon states can be more elegantly evaluated
by using the reduction formula
of the antisymmetrizer
for the three octet-baryon states~\cite{Toki:1982ec,Nakamoto:2016dmr},
\begin{align}
\mathcal{A}=&\left[
1 - 9 (P_{36}+P_{39}+P_{69}) + 27 (P_{369}+P_{396})
+54 (P_{25}\,P_{39}+P_{35}\,P_{69}+P_{38}\,P_{69})
\right]\mathcal{A}_B
\nonumber\\
&-216 P_{25}\,P_{38}\,P_{69}
\ ,\label{Eq:Toki}\\
\mathcal{A}_B=&
\sum_{\mathcal{P}} (-1)^{\pi(\mathcal{P})}\,\mathcal{P}
\ ,
\end{align}
where $P_{ij}$ denotes the exchange operator of the $i$-th and $j$-th quark, 
$P_{ijk}$ denotes the cyclic exchange operator of the $ijk$-th quarks,
and $\mathcal{P}$ is the baryon exchange operator.
Equation \eqref{Eq:Toki} takes care of 1680 permutations,
which act on $(3!)^3=216$ terms
in the product of three antisymmetrized baryon wave functions
and result in $1680\times 216=9!$ permutations.
In our categorization
the first, second, third and fourth terms
in the first line of Eq.~\eqref{Eq:Toki} correspond to 
(a1-a3), (b1-b4), (d1-d3), and (c1-c3), respectively,
and the second line corresponds to type (e).

\section{Expectation value of the KMT operator and three-baryon potential}
\label{Sec:KMT}

\subsection{KMT operator and its expectation value}

We now evaluate the KMT matrix elements
in one-, two- and three-baryon systems.
We concentrate on the $(0s)^6$ and $(0s)^9$ configurations of quarks
for two- and three-baryon systems, respectively,
which correspond to the cases where two- or three-baryons
are located at the same spatial point.

In the non-relativistic treatment of the KMT interaction \eqref{Eq:Phi},
we ignore terms which involve $\gamma_5$
and replace $\gamma_0$ with a unit matrix
in $\Phi_{ij}$,
\begin{align}
V_\mathrm{KMT}
\simeq&-2\gD \int d^3x
\,\varepsilon_{ijk}\,
u^\dagger(\bm{x}) q_i(\bm{x})\,
d^\dagger(\bm{x}) q_j(\bm{x})\,
s^\dagger(\bm{x}) q_k(\bm{x})\,
\ .
\end{align}
Then we obtain the matrix element of the KMT interaction
for the three-quark states
having a common color configuration as,
\begin{align}
&\VEV{q_1 q_2 q_3\mid{V_\KMT}\mid\, q'_1 q'_2 q'_3}
\nonumber\\
=& -2\gD\,\sum_{\{\alpha,\beta,\gamma\}}\,\int d^3x\,\varepsilon_{ijk}\,
\VEV{q_\alpha\mid u^\dagger(\bm{x}) q_i(\bm{x})\mid{q'_\alpha}}
\VEV{q_\beta \mid d^\dagger(\bm{x}) q_j(\bm{x})\mid{q'_\beta}}
\VEV{q_\gamma\mid s^\dagger(\bm{x}) q_k(\bm{x})\mid{q'_\gamma}}
\nonumber\\
=& -2\gD\,\varepsilon_{ijk}\,
\sum_{\{\alpha,\beta,\gamma\}}
\VEV{q_\alpha\mid \hat{T}^{u,i}\mid{q'_\alpha}}_\mathrm{fs}
\VEV{q_\beta\mid  \hat{T}^{d,j}\mid{q'_\beta}}_\mathrm{fs}
\VEV{q_\gamma\mid \hat{T}^{s,k}\mid{q'_\gamma}}_\mathrm{fs}
\int d^3x\,\prod_\mu
\varphi_\mu^*(\bm{x})
\varphi'_\mu(\bm{x})
\ ,
\end{align}
where the operator $\hat{T}^{i,j}$ replaces the flavor $j$ with the flavor $i$,
and $\varphi_\alpha(\bm{x})$ ($\varphi'_\alpha(\bm{x})$)
shows the spatial part of the $\alpha$-th quark wave function
in the bra (ket) state.
We
find the quantum mechanical KMT interaction operator as,
\begin{align}
V_\mathrm{KMT}
=&-2\gD \varepsilon_{ijk} \sum_{\{\alpha,\beta,\gamma\}}
	\,\hat{T}_\alpha^{u,i}\,\hat{T}_\beta^{d,j}\,\hat{T}_\gamma^{s,k}
	\delta(\bm{x}_\alpha-\bm{x}_\beta)
	\delta(\bm{x}_\beta-\bm{x}_\gamma)
\ ,\label{Eq:KMTintop}
\end{align}
The operator $\hat{T}_\alpha^{i,j}$ now acts on the $\alpha$-th quark%
.
For more generic product wave functions
with the same color configurations,
$\phi=\prod_\alpha q_\alpha$
and
$\phi'=\prod_\alpha q'_\alpha$,
the KMT matrix element is found to be
\begin{align}
\VEV{\phi\mid{V_\KMT}\mid\phi'}
=&
\sum_{\{\alpha,\beta,\gamma\}}
\VEV{q_\alpha q_\beta q_\gamma
\mid{V_\KMT}
\mid q'_\alpha q'_\beta q'_\gamma}
\prod_{i\not=\{\alpha,\beta,\gamma\}}\VEV{q_i\mid q'_i}
\nonumber\\
=&-2\gD\,
\VEV{\sigma\mid\sigma'}\,
\sum_{\{\alpha,\beta,\gamma\}}
F^\KMT_{\alpha\beta\gamma}(f,f')
R^\KMT_{\alpha\beta\gamma}(\varphi,\varphi')
\ ,\\
\VEV{\sigma\mid\sigma'}
=&\prod_\alpha \VEV{\sigma_\alpha\mid\sigma'_\alpha}
\ ,\\
F^\KMT_{\alpha\beta\gamma}(f,f')
=&
\VEV{f\mid
	\varepsilon_{ijk}
	\,\hat{T}_\alpha^{u,i}\,\hat{T}_\beta^{d,j}\,\hat{T}_\gamma^{s,k}
\mid{f'}}
\nonumber\\
=&
\delta_{u,f_\alpha}\delta_{d,f_\beta} \delta_{s,f_\gamma}
\sum_{ijk} \varepsilon_{ijk}\,
\delta_{i,f'_\alpha} \delta_{j,f'_\beta} \delta_{k,f'_\gamma}
\prod_{\mu\not=\{\alpha,\beta\,\gamma\}}\delta_{f_\mu,f'_\mu}
\ ,\\
R^\KMT_{\alpha\beta\gamma}(\varphi,\varphi')
=&
\VEV{\varphi\mid
\delta(\bm{x}_\alpha-\bm{x}_\beta)
\delta(\bm{x}_\beta-\bm{x}_\gamma)
\mid\varphi'}
\nonumber\\
=&\int d^3x
\varphi^*_\alpha(\bm{x}) \varphi^*_\beta(\bm{x}) \varphi^*_\gamma(\bm{x})
\varphi'_\alpha(\bm{x}) \varphi'_\beta(\bm{x}) \varphi'_\gamma(\bm{x})
\prod_{\mu\not=\alpha,\beta,\gamma} \VEV{\varphi_\mu\mid\varphi'_\mu}
\ .\label{Eq:R_KMT}
\end{align}
where 
$\sigma_\alpha$, $f_\alpha$ and $\varphi_\alpha$
($\sigma'_\alpha$, $f'_\alpha$ and $\varphi'_\alpha$)
are the spin, flavor and spatial wave functions
for the $\alpha$-th quark in $\phi$ ($\phi'$), respectively.

The KMT matrix elements in baryon systems are obtained as a sum of
those for the product wave functions,
\begin{align}
\mathcal{V}_\mathcal{A}\equiv
&\bra{\psi_\mathcal{A}} V_\mathrm{KMT}\ket{\psi'_\mathcal{A}}
=\bra{\psi} V_\mathrm{KMT}\ket{\mathcal{A}[\psi']}
=\sum_{I,J} c^*_I c_J \bra{\phi_I} V_\mathrm{KMT}\ket{\mathcal{A}[\phi'_J]}
\nonumber\\
=&-2\gD\sum_{I,J} 
c^*_{I} c'_{J} \sum_P C_P(\phi_I,\phi_J)\,
\VEV{\sigma_I\mid{P\sigma'_J}}\,
\sum_{\{\alpha,\beta,\gamma\}}
F^\KMT_{\alpha\beta\gamma}(f_I,Pf'_{J})
R^\KMT_{\alpha\beta\gamma}(\varphi_I,P\varphi'_{J})
\ ,
\end{align}
where
$c_I$, $f_I$ and $\varphi_I$ ($c'_J$, $f'_J$ and $\varphi'_J$ ) 
are 
the flavor-spin coefficients,
flavor configurations, and spatial wave functions of the $I$-th ($J$-th)
component
of the baryon wave function in the bra (ket) state, respectively,
and $C_P$ is the color factor.

\begin{table}
\caption{Norm and diagonal matrix elements of the KMT operator
in 
two- and three-baryon systems
of $N$ and $\Lambda$.
$2B$ and $3B$ matrix elements are evaluated
by subtracting the contribution of subsystems (see text).
For the one-baryon state,
norm is unity and the KMT diagonal matrix element is zero.
}\label{Table:KMT}
\centerline{
\begin{tabular}[t]{lcccc}
\hline
\hline
Baryon(s)	& $\mathcal{N}_\mathcal{A}$
		& $\TKMT_\mathcal{A}$
		& $\TKMT$ 
		& $\TKMT_{nB} (n=2,3)$ \\
\hline
$(NN)_{(S,T)=(0,1),(1,0)}$	& $10/9$ & $0$	  & $0$    & $0$ \\ 
$N_\ua\Lup,N_\da\Ldw$
				& 1      & $20/3$ & $20/3$ & $20/3$ \\
$N_\ua\Ldw, N_\da\Lup$		& 1      & $10/3$ & $10/3$ & $10/3$ \\
$(\Lambda\Lambda)_{S=0}$	& 1      & $18/3$ & $18/3$ & $18/3$ \\
\hline
$(NNN)_{(S,T)=(1/2,1/2)}$
				& $100/81$	& $0$        & $0$  & $0$\\ 
%
$\nup\ndw\Lambda,\pup\pdw\Lambda$
				& $25/27$	& $350/27$   & $14$ & $12/3$\\
$\nup\pup\Lup,\ndw\pdw\Ldw$	& $25/27$	& $750/27$   & $30$ & $50/3$\\
$\nup\pup\Ldw,\ndw\pdw\Lup$	& $25/27$	& $250/27$   & $10$ & $10/3$\\
$\nup\pdw\Lambda,\ndw\pup\Lambda$
				& $25/27$	& $425/27$   & $17$ & $21/3$\\
$N\Lup\Ldw$			& $45/54$	& $1035/54$  & $23$ & $21/3$\\
\hline
\hline
\end{tabular}
}
\end{table}

In Table \ref{Table:KMT}, we show the norm and the diagonal matrix elements of the KMT operator,
\begin{align}
\hat{\TKMT}^\KMT
=&
	\sum_{\{\alpha,\beta,\gamma\}}
	\varepsilon_{ijk}
	\,\hat{T}_\alpha^{u,i}\,\hat{T}_\beta^{d,j}\,\hat{T}_\gamma^{s,k}
\ ,\label{Eq:KMToperator}\\
\TKMT_\mathcal{A}
\equiv&
\VEV{\psi_\mathcal{A}\mid\hat{\TKMT}^\KMT\mid\psi_\mathcal{A}}
\nonumber\\
=&\sum_{I,J} 
c^*_{I} c'_{J} \sum_P C_P(\phi_I,\phi_J)\,
\VEV{\sigma_I\mid{P\sigma'}_J}\,
\sum_{\{\alpha,\beta,\gamma\}}
F^\KMT_{\alpha\beta\gamma}(f_I,Pf'_{J})
\ ,
\end{align}
in baryons sitting at $\bm{x}=0$.
Spatial wave functions are assumed to be the same for all quarks,
then the norm becomes zero when the two baryons have the same flavors and spins.
For one-baryon states, the diagonal matrix elements of the KMT operator are found to be zero.

We have summed up the contributions of all the quark permutations,
$3!=6$, $6!=720$ and $9!=362880$ for one-, two-, and three-baryon(s),
respectively.
We note that $F_{\alpha\beta\gamma}^\KMT$ is an integer,
$C_P$ is a multiple of $1/9$
and 
$c^*_I c'_J$
is a multiple of $(1/12)^{n-n_\Lambda}\times (1/2)^{n_\Lambda}$
for the diagonal matrix elements in
$n$ baryon systems with $n_\Lambda$ being the number of $\Lambda$.
Thus $\mathcal{N}_\mathcal{A}$ and $\TKMT_\mathcal{A}$
are rational numbers.
The genuine two- and three-baryon part of the matrix elements
are evaluated by subtracting the contribution in subsystems.
For example, for $\nup\ndw\Lup$ three-baryon systems,
we subtract the $N\Lambda$ contributions,
\begin{align}
\TKMT_{3B}(\nup\ndw\Lup)
=\TKMT(\nup\ndw\Lup)
-\TKMT(\nup\Lup)
-\TKMT(\ndw\Lup)
=4
\ ,
\end{align}
where $\TKMT=\TKMT_\mathcal{A}/\mathcal{N}_\mathcal{A}$.

We find that the expectation values of the KMT operator
take positive values of $3-20$ for $2B$ and $3B$ systems including hyperons.
Thus the KMT interaction is confirmed to generate a $3B$ repulsive potential
when we have hyperons in $3B$ systems.

The expectation value of the KMT operator, $\TKMT_{2B}$ and
$\TKMT_{3B}$,
is found to be spin-dependent:
It is larger for larger spin states, for instance,
$\TKMT_{2B}((N\Lambda)_{S=1})>\TKMT_{2B}((N\Lambda)_{S=0})$
and
$\TKMT_{3B}((NN\Lambda)_{S=3/2})>\TKMT_{3B}((NN\Lambda)_{S=1/2})$.
Since the KMT operator does not change the quark spin
in the nonrelativistic treatment,
more quark pairs have the same spin in the bra and ket
and larger matrix element $\TKMT_\Anti$ will appear
in larger spin states.
We note that the matrix element in the spin quartet state,
$\TKMT_{3B}((pn\Lambda)_{S=3/2})$, is much larger than others.

\subsection{Baryon potentials from the KMT interaction}

We evaluate the 3$B$ potential from the KMT interaction
(KMT-3B potential)
as the expectation value of the 3$B$ part of the KMT interaction operator
\eqref{Eq:KMTintop}
in three-baryons located
at $\bold{R}_1$, $\bold{R}_2$ and $\bold{R}_3$,\footnote{Strictly speaking,
this is the expectation value of the potential
and not the potential itself. As long as the extension of the baryon wave
function is large enough compared with the intrinsic extension of the quark
wave function, however, the present treatment gives a good estimate
of the potential.}
\begin{align}
V^\KMT_{3B}(\bold{R}_1,\bold{R}_2,\bold{R}_3)
=&\left.\frac{\mathcal{V}_\mathcal{A}(\bold{R}_1,\bold{R}_2,\bold{R}_3)}
{\mathcal{N}_\mathcal{A}(\bold{R}_1,\bold{R}_2,\bold{R}_3)}\right|_{3B}
\ ,
\end{align}
where $\left.\mathcal{V}_\Anti/\Norm_\Anti\right|_{3B}$ denotes
the 3$B$ potential part of the expectation value.
The spatial part of the intrinsic baryon wave function
is assumed to be $(0s)^3$;
all quarks are assumed to be in the $s$-wave state of the harmonic
oscillator potential and to have the same spatial extension,
\begin{align}
\varphi_{\bm{R}}(\bm{x})=\left(\frac{2\nu}{\pi}\right)^{3/4}\,
\exp(-\nu (\bm{x}-\bm{R})^2)
\ ,
\end{align}
where $\bm{R}$ is the position of the baryon
and the size parameter $\nu$ is related with the size of the quark
wave functions
$b$ as $\nu=1/(2b^2)$.
The strength of the KMT matrix element 
for three baryons located at the same position ($\bold{R}_{1,2,3}=0$)
reads
\begin{align}
V^\KMT_{3B}(\bold{R}_{1,2,3}=0)
=&-2\gD
	\TKMT_{3B}
	\int d^3x \varphi_0(\bm{x})^6 
= \frac{-2\gD}{(\sqrt{3}\pi b^2 )^3}\,\TKMT_{3B} 
= V_0\,\TKMT_{3B} 
\ .
\end{align}
By using the parameters,
$\gD\Lambda^5=-9.29$ and $\Lambda=631.4~\MeV$~\cite{Hatsuda:1994pi},
$b=0.6~\fm$~\cite{Oka:1981rj} or $b=0.5562~\fm$~\cite{Fujiwara:2006yh},
we find
\begin{align}
V_0
\equiv \frac{-2\gD}{(\sqrt{3}\pi b^2)^3}
=\frac{-2\gD\Lambda^5}{(\sqrt{3}\pi b^2\Lambda^2)^3}\,\Lambda
=\begin{cases}
1.45~\MeV &(b=0.6~\fm)\ ,\\
2.29~\MeV &(b=0.5562~\fm)\ .\\
\end{cases}
\label{Eq:V0}
\end{align}
If we take $\gD\Lambda^5=-12.36$ and $\Lambda=602.3~\MeV$~\cite{Rehberg:1995kh},
$V_0$ becomes about $1.68$ times larger than the values (\ref{Eq:V0})
because $V_0$ is linearly proportional to $\gD$.
At finite $\bold{R}_{1,2,3}$,
the spatial matrix element of the KMT operator is 
given as
\begin{align}
R^\KMT \simeq &
\int d^3x 
\varphi^*_{\bm{R}_a}(\bm{x})
\varphi^*_{\bm{R}_b}(\bm{x})
\varphi^*_{\bm{R}_c}(\bm{x})
\varphi_{\bm{R}_d}(\bm{x})
\varphi_{\bm{R}_e}(\bm{x})
\varphi_{\bm{R}_f}(\bm{x})
\nonumber\\
=&\frac{1}{(\sqrt{3}\pi{b^2})^3}
\exp\left[
-\frac{\nu}{6}\sum_{i<j,i,j=a\sim f} \bm{R}_{ij}^2
\right]
\ ,
\end{align}
where $\bm{R}_{ij}=\bm{R}_i-\bm{R}_j$
and $\bm{R}_i (i=a\sim f)$ is one of $\bm{R}_1$, $\bm{R}_2$ and $\bm{R}_3$.
When the antisymmetrization effects on the spatial wave function are ignored,
other spatial matrix elements
($\prod \VEV{\varphi_\mu\mid\varphi'_\mu}$ in Eq.~\eqref{Eq:R_KMT})
cancels with those from the norm.\footnote{For a more serious discussion,
we need to evaluate the spatial factor
$R^\KMT_{\alpha\beta\gamma}$ in Eq.~\eqref{Eq:R_KMT}
including the overlaps of other quarks, 
as well as those in the norm.
In addition, the KMT matrix element can be finite
also in those configurations having $\TKMT=0$ at zero distance.
For example, the $\nup\nup\Lup$ configuration has a zero norm
and a zero KMT matrix element due to the Pauli blocking
when three baryons are located at the same point,
but will have a finite norm and a finite KMT matrix element
at finite distances.
These are beyond the scope of this paper,
and we concentrate on the potentials
in $NN\Lambda$ and $N\Lambda\Lambda$ channels
having finite KMT matrix elements at zero distance.}
This prescription provides correct results
when the baryons are separated enough
or three-baryons are located at the same spatial point.
The KMT-3B potential is then given as
\begin{align}
V^\KMT_{3B}(\bm{R}_1,\bm{R}_2,\bm{R}_3)
\simeq & V_0 \TKMT_{3B}
\,\exp\left[-\frac{2\nu}{3}(\bm{R}_{12}^2+\bm{R}_{23}^2+\bm{R}_{31}^2)\right]
\ .
\end{align}



\begin{figure}
\centerline{\includegraphics[width=7.5cm]{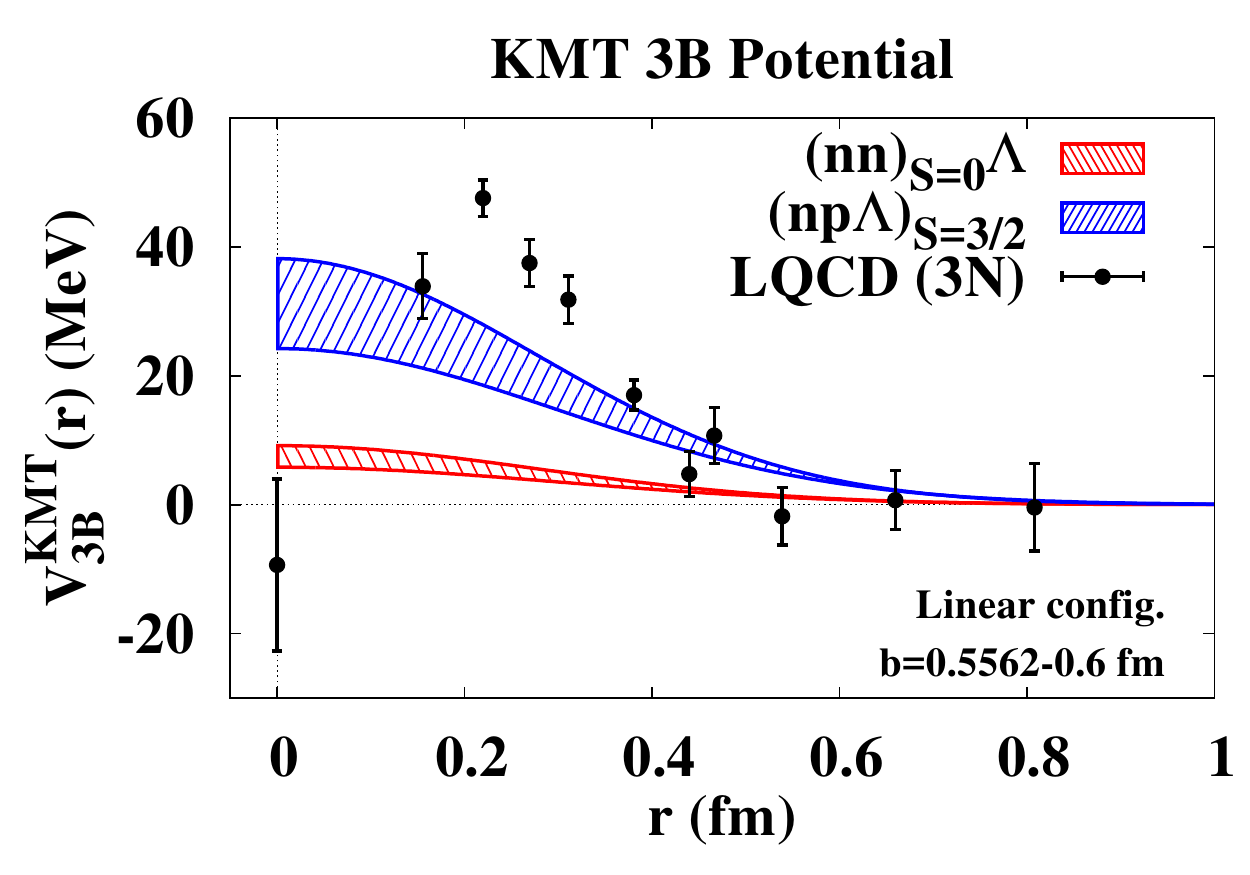}}
\caption{Three-baryon potential in $(nn)_{S=0}\Lambda$
and $(np\Lambda)_{S=3/2}$ channels generated by the KMT interaction.
Shaded area shows the uncertainties from the size parameter.
Filled circles show the results of the $3N$ potential
obtained in the lattice QCD simulation~\cite{Doi:2011gq}.
}
\label{Fig:V3}
\end{figure}

In Fig.~\ref{Fig:V3}, we show the KMT-3B potential
$V^\mathrm{KMT}_\mathrm{3B}$
for the $NN\Lambda$ system
in $(nn)_{S=0}\Lambda$ and $(np\Lambda)_{S=3/2}$ channels
with a linear configuration in which three baryons are located at
$\bold{R}_1=(-r,0,0)$,
$\bold{R}_2=(0,0,0)$
and $\bold{R}_3=(r,0,0)$.
We adopt $V_0$ values evaluated in Eq.~\eqref{Eq:V0}.
The KMT-$NN\Lambda$ potential at $r=0$ has the height of
5.8 (9.2) MeV in $(nn)_{S=0}\Lambda$ channel
and
24.2 (38.2) MeV in $(np\Lambda)_{S=3/2}$ channel
for $b=0.6 (0.5562)~\fm$, respectively,
and the potential range is determined by the baryon size.

We compare the KMT-$NN\Lambda$ potential with the 3$N$ potential
obtained by using the lattice QCD simulation (LQCD-3N potential)~\cite{Doi:2011gq}.
It is interesting to find that the KMT-$NN\Lambda$ potential and the
LQCD-3N potential have the same order of strengths and ranges, while the
system is different and they do not need to agree. One of the
differences is the short range behavior; the KMT-$NN\Lambda$ potential
has the peak at $r=0$, whereas the LQCD-3N potential is suppressed at $r=0$.

\subsection{KMT-3B potential energy in nuclear matter}

We shall now evaluate the KMT-3B potential energy per baryon in nuclear matter.
We assume that the density of each baryons is constant,
then we get the energy per baryon $W^\KMT_{3B}$
from the KMT-3B potential as
\begin{align}
W^\KMT_{3B} =& \frac{1}{\rhoB\Omega}\int d^3R_1\, d^3R_2\, d^3R_3\,
\frac{1}{3!}\sum_{B_1,B_2,B_3}\,
\rho(B_1)\,\rho(B_2)\,\rho(B_3)\,
V_{3B}^\KMT (B_1B_2B_3;\bm{R}_1,\bm{R}_2,\bm{R}_3)
\nonumber\\
=& \frac{W_0}{3!}\,\left(\frac{\rhoB}{\rho_0}\right)^2\,
\sum_{B_1,B_2,B_3}\,\frac{\rho(B_1)\rho(B_2)\rho(B_3)}{\rhoB^3}
\TKMT_{3B}(B_1B_2B_3)
\ ,\\
W_0=&
V_0\,\left(\sqrt{3}\pi{b^2}\right)^3\,\rho_0^2
=-2\gD \rho_0^2
\simeq 0.28~\MeV
\ .
\end{align}
where $\rhoB$ is the baryon density and $\Omega$ denotes the spatial volume.
It should be noted that $W_0$ is independent of the baryon size $b$.
The KMT-3B potential energy per baryon in $n\Lambda$ matter is given as
\begin{align}
W_{n\Lambda}^\KMT(\rhoB,Y_\Lambda)
=W_0 \left(\frac{\rhoB}{\rho_0}\right)^2\left[
\frac12\,Y_n^2\,Y_\Lambda\,\TTKMT_{3B}(NN\Lambda)
+\frac12\,Y_n\,Y_\Lambda^2\,\TTKMT_{3B}(N\Lambda\Lambda)
\right]\ ,
\end{align}
where $Y_n$ and $Y_\Lambda$ represent the density fractions
of $n$ and $\Lambda$,
$Y_n=\rho_n/\rhoB$ and $Y_\Lambda=\rho_\Lambda/\rhoB$, respectively,
and $\TTKMT_{3B}(NN\Lambda)$ and $\TTKMT_{3B}(N\Lambda\Lambda)$
are the spin-averaged KMT matrix elements,
\begin{align}
\TTKMT_{3B}(NN\Lambda)
=&\frac18\,\sum_{\sigma_1,\sigma_2,\sigma_3}
\TKMT_{3B}(n_{\sigma_1}n_{\sigma_2}\Lambda_{\sigma_3})
=\frac12\,\TKMT_{3B}((nn)_{S=0}\Lambda)=2
\ ,\\
\TTKMT_{3B}(N\Lambda\Lambda)
=&\frac18\,\sum_{\sigma_1,\sigma_2,\sigma_3}
\TKMT_{3B}(N_{\sigma_1}\Lambda_{\sigma_2}\Lambda_{\sigma_3})
=\frac12\,\TKMT_{3B}(N(\Lambda\Lambda)_{S=0})=\frac{7}{2}
\ ,
\end{align}
where $(S,T)=(0,1)$ pair is taken for $NN\Lambda$.
Note that $Y_n+Y_\Lambda=1$ in $n\Lambda$ matter.
For $np\Lambda$ matter,
we find
\begin{align}
W_{pn\Lambda}^\KMT(\rhoB,Y_\Lambda)
=W_0 \left(\frac{\rhoB}{\rho_0}\right)^2&\left[
\frac12\,(Y_n^2+Y_p^2)\,Y_\Lambda\,\TTKMT_{3B}(NN\Lambda)
+\frac12\,(Y_n+Y_p)\,Y_\Lambda^2\,\TTKMT_{3B}(N\Lambda\Lambda)
\right.
\nonumber\\
&\left.
+Y_n\,Y_p\,Y_\Lambda\,\TTKMT_{3B}(np\Lambda)
\right]\ ,
\end{align}
where $\TTKMT_{3B}(np\Lambda)$
is the spin-averaged KMT matrix element in the $np\Lambda$ system,
\begin{align}
\TTKMT_{3B}(np\Lambda)=&
\frac14\large[\TKMT_{3B}((np\Lambda)_{S=3/2})
+2\TKMT_{3B}(\nup\pdw\Lambda)
+\TKMT_{3B}(\nup\pup\Ldw)
\large]
=
\frac{17}{2}
\ .
\end{align}
Note that $Y_n+Y_p+Y_\Lambda=1$ in the $np\Lambda$ matter.

\begin{figure}
\centerline{%
\includegraphics[width=7.5cm]{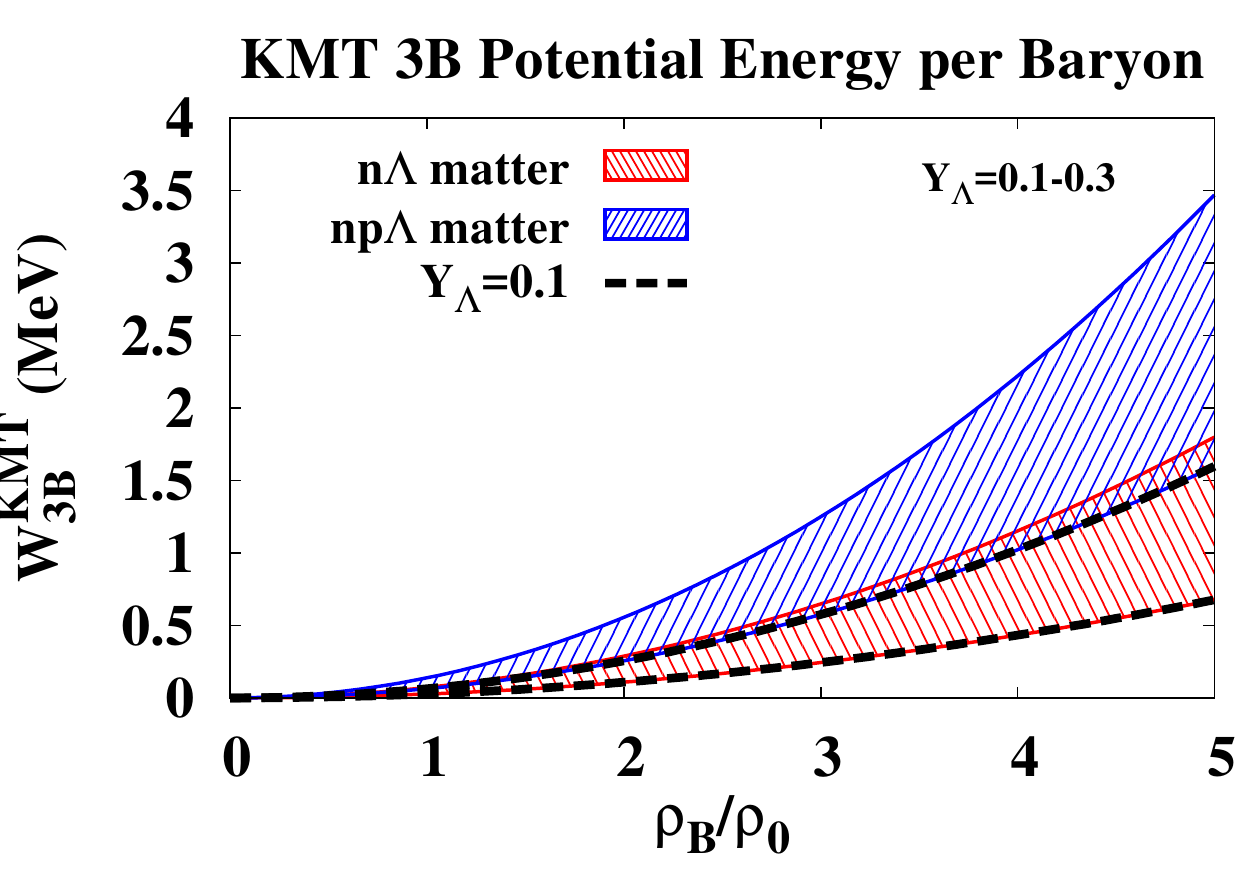}%
\includegraphics[width=7.5cm]{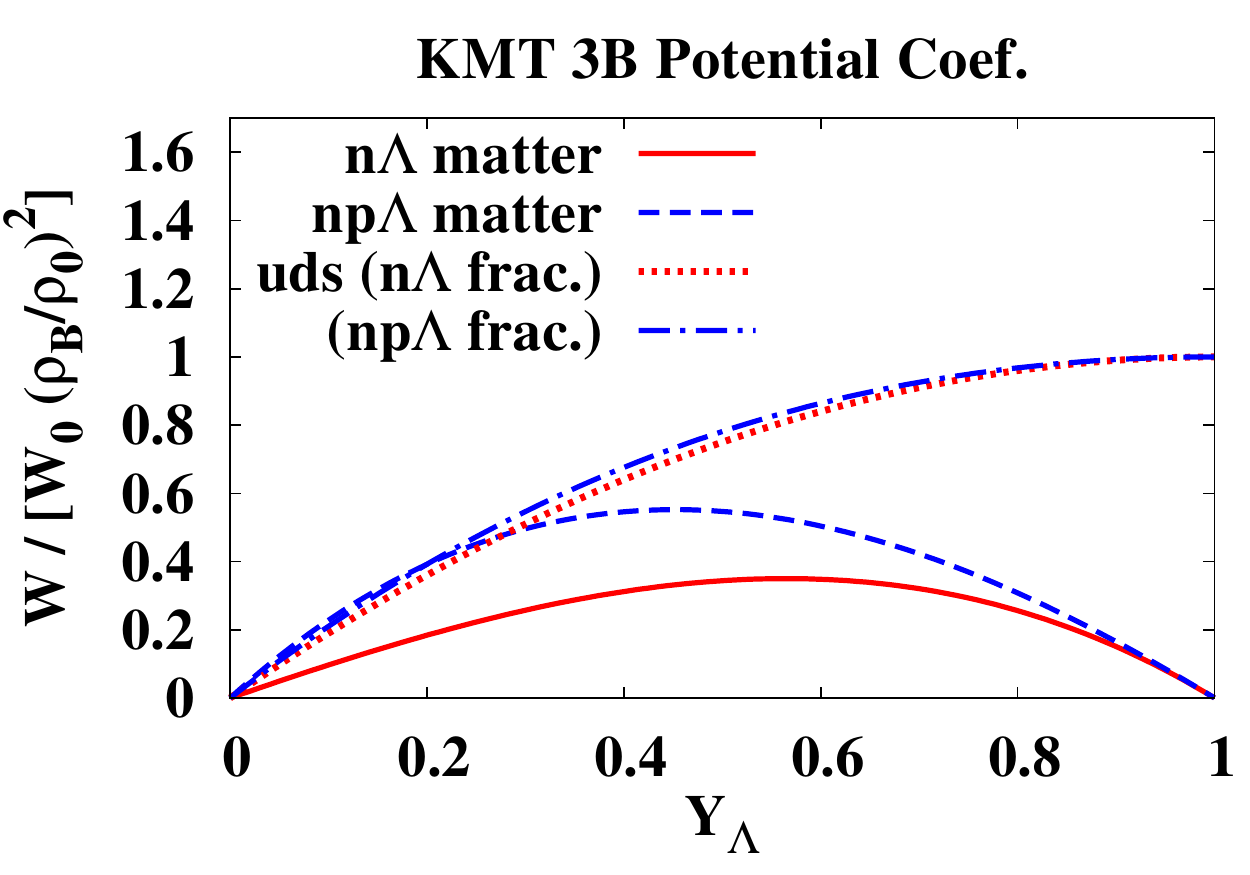}%
}
\caption{KMT-3B potential energy per baryon in $n\Lambda$ and $np\Lambda$ 
matter (left),
and the coefficients of the potential energy per baryon,
$W^\KMT/[W_0\,(\rhoB/\rho_0)^2]$
in $n\Lambda$, $np\Lambda$ and quark matter (right).
}
\label{Fig:W3}
\end{figure}

In the left panel of Fig.~\ref{Fig:W3},
we show the KMT-3B potential energy per baryon
in $n\Lambda$ and $np\Lambda$ matter.
Each shaded area corresponds to $0.1 \leq Y_\Lambda \leq 0.3$
and neutron and proton fractions are taken to be
$Y_n=1-Y_\Lambda$ and $Y_n=Y_p=(1-Y_\Lambda)/2$
for $n\Lambda$ and $np\Lambda$ matter, respectively.
The KMT-3B potential energy per baryon amounts to be
$W^\mathrm{KMT}_{3B}=0.027$ and $0.064~\MeV$
at $(\rhoB,Y_\Lambda)=(\rho_0,0.1)$
in $n\Lambda$ and $np\Lambda$ matter, respectively.
At higher density and larger $\Lambda$ fraction,
the KMT-3B potential is found to reduce the nuclear symmetry energy slightly;
$W^\mathrm{KMT}_{3B}=0.65$ and $1.25~\MeV$
at $(\rhoB,Y_\Lambda)=(3\rho_0,0.3)$
in $n\Lambda$ and $np\Lambda$ matter, respectively,
then the KMT-3B potential reduces the difference
of the energy per baryon
in neutron matter with $\Lambda$ admixture ($n\Lambda$ matter)
and
in symmetric nuclear matter with $\Lambda$ admixture ($np\Lambda$ matter)
by about
$0.6~\MeV$.

It may be instructive to compare the above potential energy
with that in quark matter having the same $uds$ quark fractions,
\begin{align}
&W_{uds}^\KMT(\rhoB)
=W_0 \left(\frac{\rhoB}{\rho_0}\right)^2
\, Y_u\,Y_d\,Y_s
\ ,\\
&Y_u=\rho_u/\rhoB=2Y_p+Y_n+Y_\Lambda\ ,\quad
Y_d=Y_p+2Y_n+Y_\Lambda\ ,\quad
Y_s=Y_\Lambda\ .
\end{align}
It should be noted that only the residual interaction contribution
is considered here,
and other contributions coming from the condensates are not counted.
In the right panel of Fig.~\ref{Fig:W3}, we compare the coefficients
of the potential energy per baryon,
$W^\KMT/[W_0\,(\rhoB/\rho_0)^2]$
in $n\Lambda$, $np\Lambda$ and quark matter.
At small strange quark fraction (small $Y_\Lambda$),
we note that the quark matter estimate agrees with the baryonic matter estimate
in $np\Lambda$ matter,
while baryonic
matter estimate 
results in
weaker KMT-3B repulsion in
$n\Lambda$ matter.

Finally, we discuss the KMT-3B potential contribution to the $\Lambda$
single particle energy,
\begin{align}
U_\Lambda
=\frac{\partial\left(\rhoB W\right)}{\partial \rho_\Lambda}
\ .
\end{align}
In $n\Lambda$ and $pn\Lambda$ matter, we find
\begin{align}
U_{\Lambda(n\Lambda)}^\KMT(\rhoB,Y_\Lambda)
=&W_0 \left(\frac{\rhoB}{\rho_0}\right)^2\left[
\frac12\,Y_n^2\,\TTKMT_{3B}(NN\Lambda)
+Y_n\,Y_\Lambda\,\TTKMT_{3B}(N\Lambda\Lambda)
\right]\ ,
\\
U_{\Lambda(pn\Lambda)}^\KMT(\rhoB,Y_\Lambda)
=&W_0 \left(\frac{\rhoB}{\rho_0}\right)^2\left[
\frac12\,(Y_n^2+Y_p^2)\,\TTKMT_{3B}(NN\Lambda)
+(Y_n+Y_p)\,Y_\Lambda\,\TTKMT_{3B}(N\Lambda\Lambda)
\right.
\nonumber\\
&\hspace*{2cm}\left.
+Y_n\,Y_p\,\TTKMT_{3B}(np\Lambda)
\right]\ ,
\end{align}

\begin{figure}
\centerline{\includegraphics[width=7.5cm]{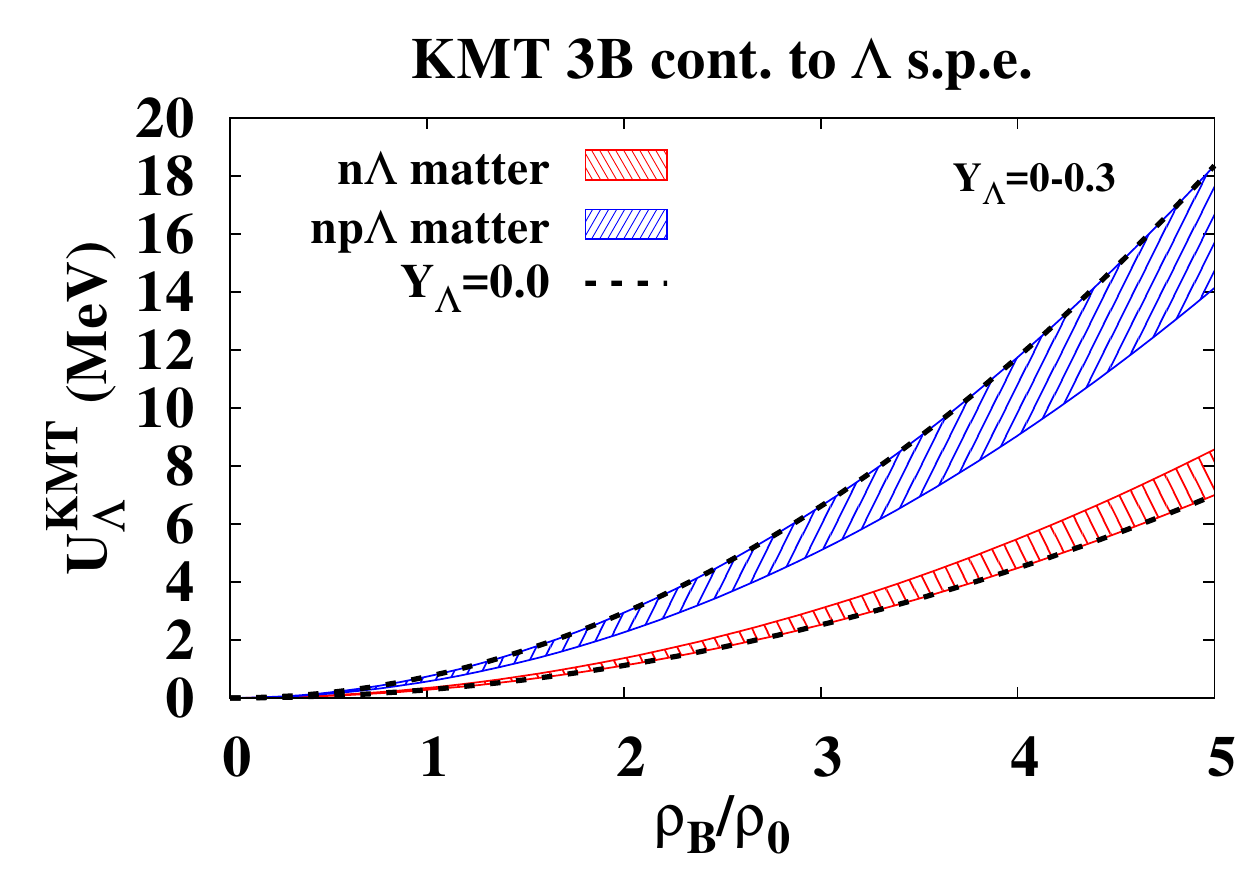}}
\caption{KMT-3B potential contribution to $\Lambda$ single particle energy.}
\label{Fig:U3}
\end{figure}

In Fig.~\ref{Fig:U3}, we show the $\Lambda$ single particle potential
in $n\Lambda$ and $pn\Lambda$ matter.
The neutron and proton fractions in $np\Lambda$ matter is again taken to be
$Y_n=Y_p=(1-Y_\Lambda)/2$.
The average $pn\Lambda$ matrix element is larger than that of $nn\Lambda$,
since we have more combinations of $uds$ in $pn\Lambda$ matter.
As a result, $\Lambda$ feels more repulsion in symmetric matter
than in pure neutron matter.
The KMT-3B potential contribution to the $\Lambda$ single particle energy
is found to be 
$U_{\Lambda(n\Lambda)}^\KMT\simeq 0.28~\MeV$
and
$U_{\Lambda(pn\Lambda)}^\KMT\simeq 0.73~\MeV$
at $\rhoB=\rho_0$
in neutron matter and symmetric nuclear matter
($n\Lambda$ and $pn\Lambda$ matter at $Y_\Lambda=0$), respectively.
These contributions are not large,
but
would be measurable in high-resolution experiments
on the $\Lambda$ separation energy of hyperisotopes
such as $^{40}_\Lambda\mathrm{K}$ and $^{48}_\Lambda\mathrm{K}$
\cite{JLabHypernuclearCollaboration}.


\section{Summary}
\label{sec:summary}

We have evaluated the expectation value
of the determinant interaction of quarks,
the Kobayashi-Maskawa~\cite{Kobayashi:1970ji,Kobayashi:1971qz}
and
't Hooft~\cite{'tHooft:1976fv,'tHooft:1986nc} (KMT) interaction,
in three-baryon systems,
and have discussed the three-baryon potential from the KMT interaction.
The KMT vertex gives rise to such three-quark interaction
that all the $u$, $d$ and $s$ quarks need to participate.
Then the KMT interaction is expected to generate potentials
among three baryons which include at least one hyperon.
The KMT interaction is responsible to the $U(1)_A$ anomaly,
and its strength is determined 
from the $\eta'$ mass~\cite{Hatsuda:1994pi,Rehberg:1995kh}.
The negative sign of the strength parameter in the Lagrangian
generates a repulsive three-quark interaction.
Repulsive potential among three-baryons including hyperons
may help to solve the hyperon puzzle
of two-solar-mass neutron
stars~\cite{Demorest:2010bx,Antoniadis:2013pzd}.

The expectation value of the KMT operator (Eq.~\eqref{Eq:KMToperator})
is obtained for the two- and three-baryon (2$B$ and 3$B$) states
consisting of nucleons and $\Lambda$ baryons,
where baryons are assumed to be located at the same spatial point.
We have adopted nonrelativistic $(0s)^3$ wave functions for octet baryons,
where a common spatial wave function is assumed for all quarks.
2$B$ and 3$B$ states are given as the product of baryon wave functions
antisymmetrized under the quark exchanges.
The KMT operator is found to take positive expectation values
of $3-20$
for  $2B$ and $3B$ systems consisting of $N$ and at least one
$\Lambda$. 
Thus the KMT interaction is confirmed to generate $3B$ repulsive potential
in $NN\Lambda$ and $N\Lambda\Lambda$ systems.
The expectation value of the KMT operator, $\TKMT_{2B}$ and $\TKMT_{3B}$,
is found to be larger for the larger spin state,
$\TKMT_{2B}((N\Lambda)_{S=1})>\TKMT_{2B}((N\Lambda)_{S=0})$
and
$\TKMT_{3B}((NN\Lambda)_{S=3/2})>\TKMT_{3B}((NN\Lambda)_{S=1/2})$.
Since the KMT operator does not change the quark spin
in the nonrelativistic treatment,
there can be more quark pairs having the same spin
in the bra and ket of larger spin states and the expectation value
of the KMT operator tends to be larger.
We also note that the matrix element in the spin quartet state,
$\TKMT_{3B}((pn\Lambda)_{S=3/2}$, is much larger than others.

We have evaluated the $3B$ potential from the KMT interaction
(KMT-3B potential).
The strength of the 3$B$ potential is calculated
by using the harmonic oscillator wave function of quarks
adopted in the quark cluster model analyses of 2$B$
potentials~\cite{Oka:1981rj,Fujiwara:2006yh}
and the strength of the KMT interaction obtained
from the $\eta'$ mass analyses~\cite{Hatsuda:1994pi}.
The obtained $NN\Lambda$ potential from the KMT interaction has the height of
$5.8-9.2$ MeV in $(nn)_{S=0}\Lambda$ channel
and
$24.2-38.2$ MeV in $(np\Lambda)_{S=3/2}$ channel
at zero distance,
and the potential range is determined by the baryon size.
The strength and the range of the $3B$ potential
in the linear configuration 
are found to be similar to those from the $3N$ potential
obtained in the lattice QCD simulation~\cite{Doi:2011gq}
except the short range region.

The KMT-3B potential energy per baryon in nuclear matter $W^\KMT_{3B}$
and the KMT-3B potential contribution to the $\Lambda$ single particle potential
$U^\KMT_\Lambda$ are also estimated.
The KMT-3B contribution to the equation of state, $W^\KMT_{3B}$,
is found to be small;
$W^\KMT_{3B}\simeq 0.05~\MeV$ and $0.1~\MeV$,
for $n\Lambda$ and $np\Lambda$ matter
at $(\rhoB,Y_\Lambda)=(\rho_0, 0.1)$, respectively.
The KMT-3B contribution to the $\Lambda$ single particle potential,
$U^\KMT_\Lambda$, is not large but would be visible;
$U^\KMT_\Lambda \simeq 0.56~\MeV$ and $1.2~\MeV$
in $n\Lambda$ and $np\Lambda$ matter at $\rhoB=\rho_0$, respectively.
The difference is small, but may be detectable in high-precision experiment
on the separation energies of hyperisotopes
such as $^{40}_\Lambda\mathrm{K}$
and $^{48}_\Lambda\mathrm{K}$~\cite{JLabHypernuclearCollaboration}.
At $\rhoB=3\rho_0$, $U^\KMT_\Lambda$ amounts to
$2.5~\MeV$ and $6.6~\MeV$
in $n\Lambda$ and $np\Lambda$ matter, respectively.
In order to solve the hyperon puzzle of massive neutron stars,
we need to find the mechanism to generate additional $\Lambda$ single particle
potential $\Delta U_\Lambda \sim 100~\MeV$ at around $\rhoB=3\rho_0$,
and the KMT-3B potential explains $2-7 \%$ of the required repulsion.
Other three-quark interactions such as
the confinement potential~\cite{Takahashi:2002bw}
can also contribute to the 3$B$ potential.

\section*{Acknowledgments}
This work was supported in part by
the Grants-in-Aid for Scientific Research on Innovative Areas from MEXT
(Nos. 24105008 and 24105001),
Grants-in-Aid for Scientific Research from JSPS
(Nos.
  15K05079, 
  15H03663, 
  16K05349, 
  16K05350
).
K.M. was supported in part by the National Science Center, Poland under
Maestro Grant No.  DEC-2013/10/A/ST2/00106.
K.K. was supported in part by Grant-in-Aid No. 26-1717 from JSPS.

\appendix

\section{Expectation values of KMT operator in one-, two-, and three-baryon
systems}

\begin{figure}
\centerline{%
\includegraphics[width=10.5cm]{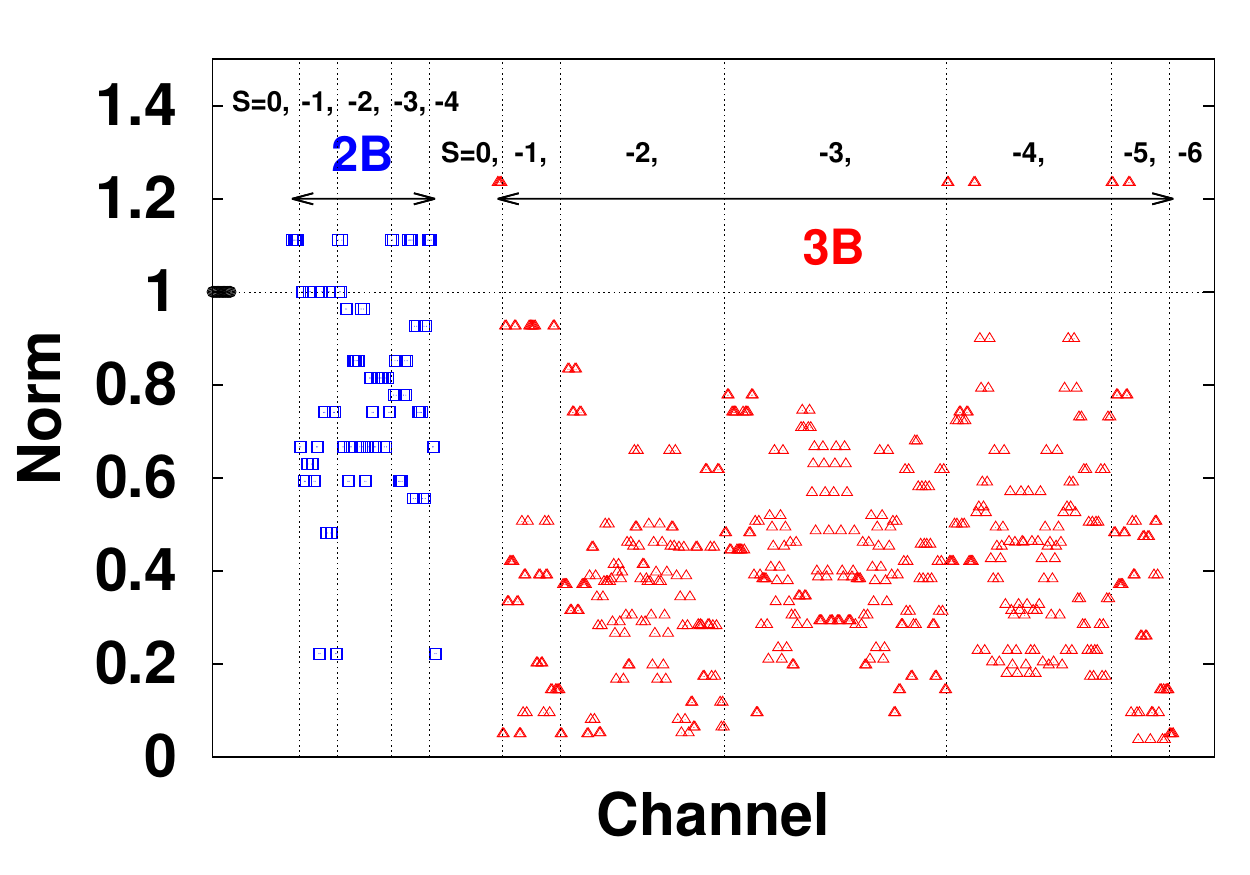}%
}
\centerline{
\includegraphics[width=10.5cm]{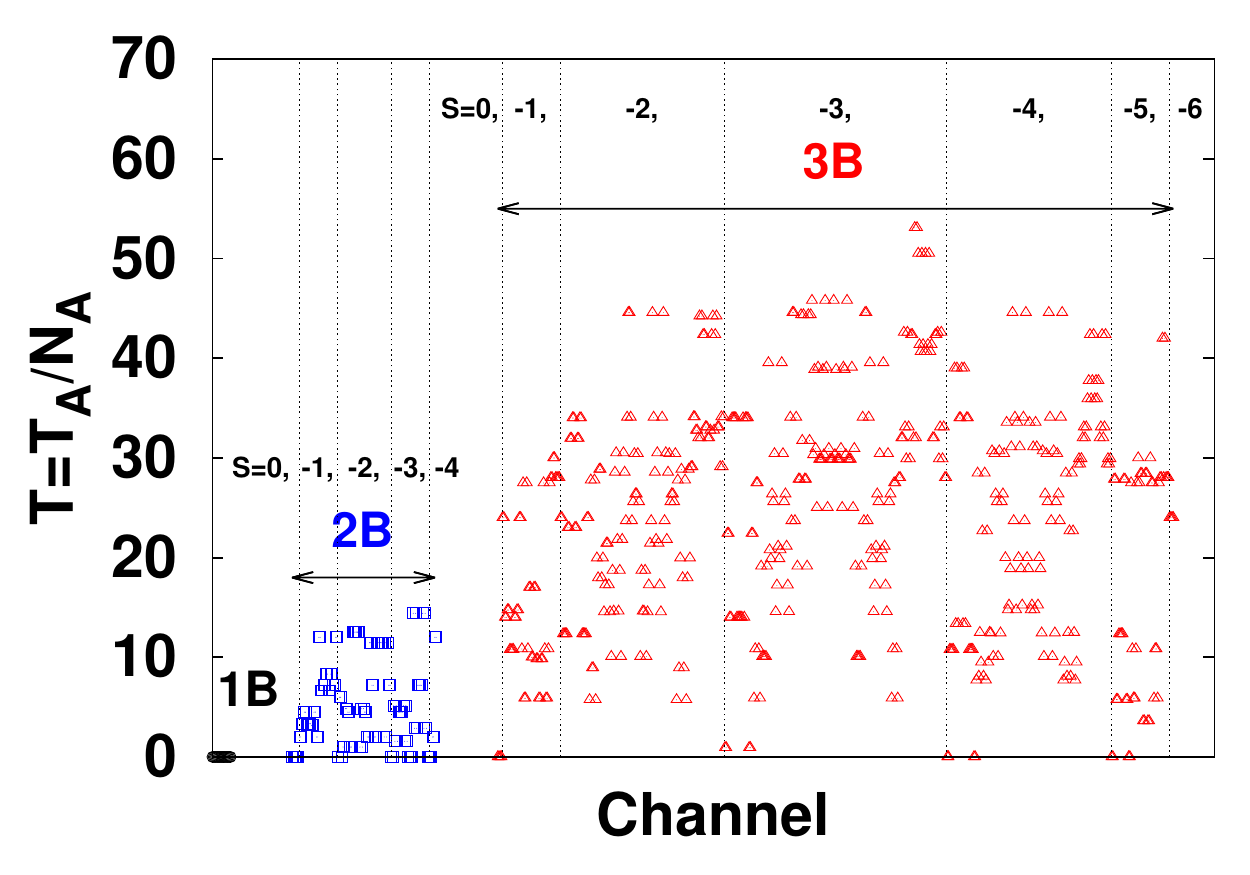}%
}
\centerline{
\includegraphics[width=10.5cm]{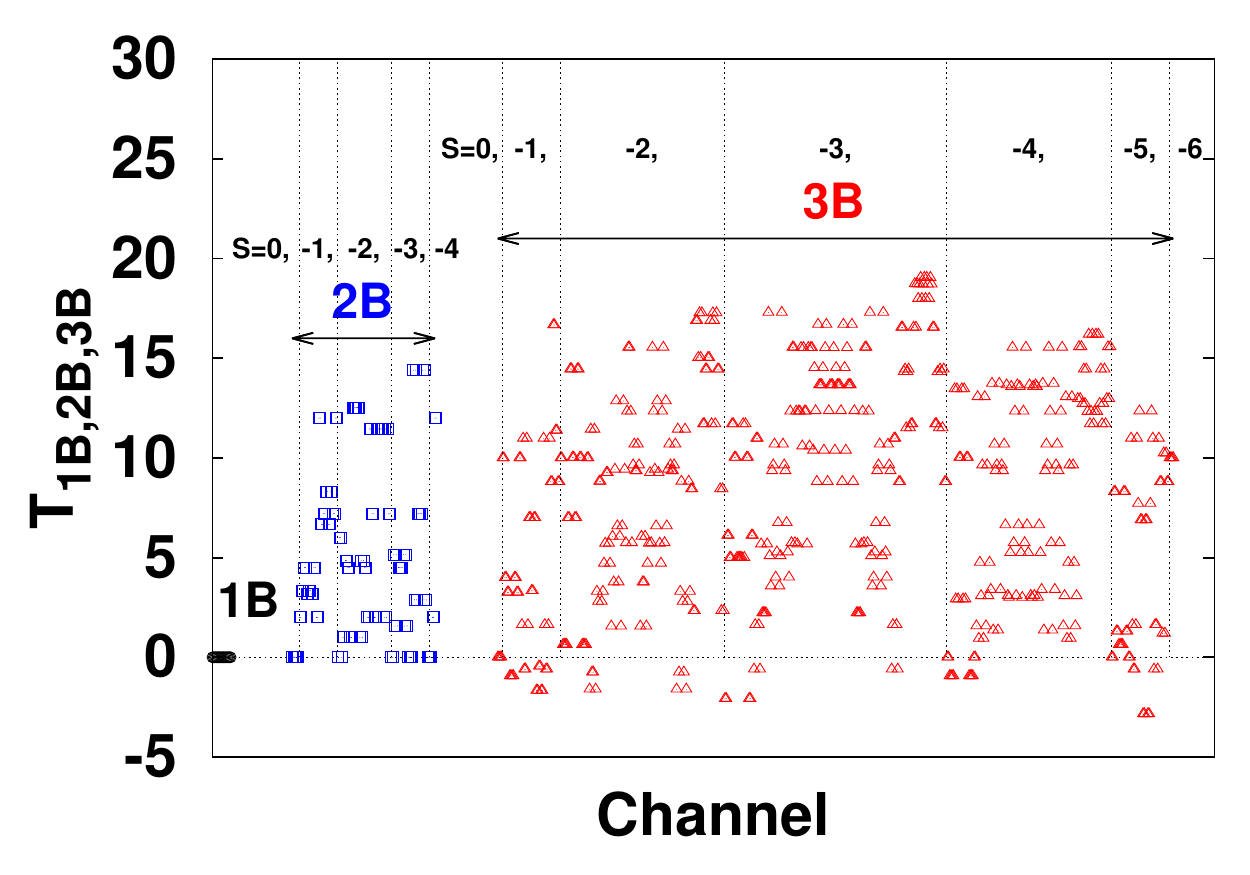}%
}
\caption{Norm (top), expectation value of the KMT operator (middle)
and its $nB$ part (bottom)
in one-, two-, and three-octet baryon states.
}\label{Fig:KMTall}
\end{figure}

We have shown in Sec.~\ref{Sec:KMT}
the expectation values of the KMT operator
in some selected channels including $N$ and $\Lambda$.
In Fig.~\ref{Fig:KMTall},
we show the results for all states consisting of
one-, two-, and three-octet baryons,
\begin{align}
\ket{\psi_\mathcal{A}}=
\ket{B_\sigma}
\ ,\quad
\frac{\mathcal{A}}{\sqrt{6!}}
\ket{B_\sigma B'_{\sigma'}}
\ ,\quad
\frac{\mathcal{A}}{\sqrt{9!}}
\ket{B_\sigma B'_{\sigma'} {B''_{\sigma''}}}
\ .
\end{align}
Channels are sorted according to strangeness quantum number.

As already discussed in Ref.~\cite{Oka:1986fr,Nakamoto:2016dmr},
the norm
takes small values in, for example, 
$(N\Sigma)_{(S,T)=(1,3/2)}$,
$(\Xi^-\Xi^0)_{S=1}$
$(\Norm_\Anti=2/9)$
and 
$(N\Sigma^0)_{S=1}$,
$(\Norm_\Anti=13/27)$, 
in 2$B$ states,\footnote{To describe
$(N\Sigma)_{(S,T)=(0,1/2)}$ state having $\Norm_\Anti=1/9$
or
$(N\Lambda-N\Sigma)_{(S,T)=(0,1/2)}$ coupled state
having $\Norm_\Anti=0, 10/9$,
off-diagonal matrix elements have to be accounted for.}
and
$\Lambda(\Xi^-\Xi^0)_{S=1}$
$(\Norm_\Anti=1/27)$, 
$(nn)_{S=0}\Sigma^-$,
$n(\Sigma^-\Sigma^-)_{S=0}$,
$(pp)_{S=0}\Sigma^+$,
$p(\Sigma^+\Sigma^+)_{S=0}$,
$(\Xi^-\Xi^-)_{S=0}\Xi^0$,
$\Xi^-(\Xi^0\Xi^0)_{S=0}$,
$(\Norm_\Anti=4/81)$, 
in 3$B$ states,
where repulsion from the quark-Pauli effects is expected.

The expectation value of the KMT operator
takes large values in, for example, 
$(\Lambda\Xi)_{S=1}$ 
$(\TKMT=72/5)$,
$\pup\Xmdw$, $\pdw\Xmup$, $\nup\Xzdw$,
$\ndw\Xzup$, $\Smup\Spdw$, $\Smdw\Spup$
$(\TKMT=288/23)$,
$(n\Sigma^-)_{S=1}$, $(p\Sigma^+)_{S=1}$, $(\Xi^-\Xi^0)_{S=1}$,
$(\TKMT=12)$,
in 2$B$ states,
and
$(\Sigma\Sigma\Sigma)_{S=3/2}$
$(\TKMT=584/11)$,
$(p\Lambda\Xi^-)_{S=3/2}$ $(n\Lambda\Xi^0)_{S=3/2}$
$(\TKMT=2374/47)$,
$\pdw\Lup\Xmup$ $\pup\Ldw\Xmdw$ $\ndw\Lup\Xzup$ $\nup\Ldw\Xzdw$
$(\TKMT=2106/46)$,
in 3$B$ states.
In some of
$NN\Sigma$, 
$N\Lambda\Sigma$, 
$N\Sigma\Xi$, 
$N\Xi\Xi$, 
$\Lambda\Sigma\Sigma$, 
$\Sigma\Xi\Xi$,
and $\Sigma\Sigma\Xi$ states,
the 3$B$ part of the expectation value is found to be negative,
while $\TKMT$ is positive.

\end{document}